\documentclass[times,twocolumn,final]{elsarticle}
\usepackage{medima}
\usepackage{framed,multirow}

\usepackage{amssymb}
\usepackage{latexsym}
\usepackage{graphicx}
\usepackage{makecell}
\usepackage{amsmath}
\usepackage{amsfonts,amssymb} 
\usepackage{caption}
\usepackage{subfigure}
\usepackage{wrapfig}

\usepackage{parskip}
\usepackage{booktabs}
\usepackage{multirow}
\usepackage{xcolor}

\usepackage{url}
\usepackage{xcolor}

\usepackage{hyperref}

\definecolor{newcolor}{rgb}{.8,.349,.1}

\begin{document}

\verso{Xuan Xu \textit{et~al.}}

% \begin{frontmatter}

\title{Histo-Diffusion: A Diffusion Super-Resolution Method for Digital Pathology with Comprehensive Quality Assessment}%
% \tnotetext[tnote1]{This is an example for title footnote coding.}

\author[1]{Xuan \snm{Xu}\corref{cor1}}
\cortext[cor1]{Corresponding author: 
  Stony Brook University}
\author[1]{Saarthak \snm{Kapse}}
\author[1]{Prateek \snm{Prasanna}}
% \fntext[fn1]{This is author footnote for second author.}
% \author[2]{Given-name3 \snm{Surname3}}
% %% Third author's email
% \ead{author3@author.com}
% \author[2]{Given-name4 \snm{Surname4}}

\address[1]{Stony Brook University}
% \address[2]{Affiliation 2, Address, City and Postal Code, Country}

\received{11 Aug 2024}
% \finalform{10 May 2013}
% \accepted{13 May 2013}
% \availableonline{15 May 2013}
% \communicated{S. Sarkar}

\begin{abstract}

Digital pathology has advanced significantly over the last decade, with Whole Slide Images (WSIs) encompassing vast amounts of data essential for accurate disease diagnosis. High-resolution WSIs are essential for precise diagnosis but technical limitations in scanning equipment and variability in slide preparation can hinder obtaining these images. Super-resolution techniques can enhance low-resolution images; while Generative Adversarial Networks (GANs) have been effective in natural image super-resolution tasks, they often struggle with histopathology due to overfitting and mode collapse. Traditional evaluation metrics fall short in assessing the complex characteristics
of histopathology images, necessitating robust histology-specific evaluation methods.

We introduce Histo-Diffusion, a novel diffusion-based method specially designed for generating and evaluating super-resolution images in digital pathology. It includes a restoration module for histopathology prior and a controllable diffusion module for generating high-quality images. We have curated two histopathology datasets and proposed a comprehensive evaluation strategy which incorporates both full-reference and no-reference metrics to thoroughly assess the quality of digital pathology images.

Comparative analyses on multiple datasets with state-of-the-art methods reveal that Histo-Diffusion outperforms GANs. Our method offers a versatile solution for histopathology image super-resolution, capable of handling multi-resolution generation from varied input sizes, providing valuable support in diagnostic processes.
\end{abstract}

\begin{keyword}
% \MSC 41A05\sep 41A10\sep 65D05\sep 65D17
%% Keywords
\KWD Image Super Resolution\sep Digital Pathology\sep Diffusion models
\end{keyword}

\maketitle

\section{Introduction}

In recent years, the field of digital pathology has seen a marked increase in interest, driven primarily by the increased availability of Whole Slide Images captured through advanced scanners. 
% Central to this field is the use of Hematoxylin and Eosin (H\&E) stain, a foundational tissue stain in histology, widely regarded as the gold standard in medical diagnostics. For instance, when examining a biopsy from a potentially cancerous lesion, pathologists typically rely on slides treated with H\&E stain.
WSIs, known for their considerable data size, often amounting to several gigabytes per slide, contain tens of thousands of nuclei and other primitives essential for detailed analysis necessary in disease diagnosis. Consequently, high-resolution WSIs are crucial for enabling precise visualization, improving diagnostic accuracy, and facilitating automated analysis and accurate measurements.
\begin{figure*}[!t]
\centering
\includegraphics[scale=.9]{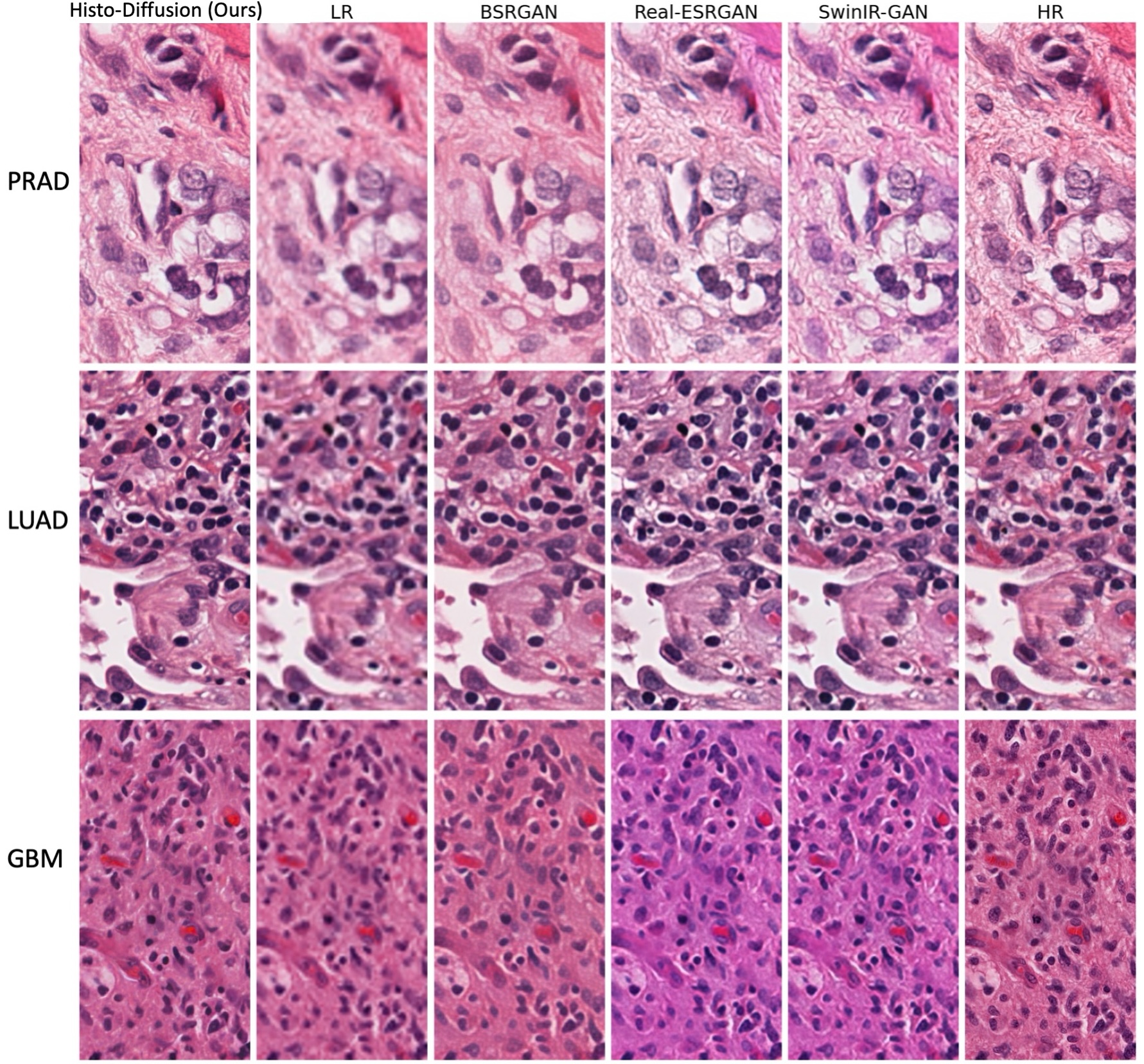}
\caption{Comparisons of state-of-the-art methods and Histo-Diffusion.}
\label{teaser0}
\end{figure*}
\begin{figure*}[!t]
\centering
\includegraphics[scale=.20]{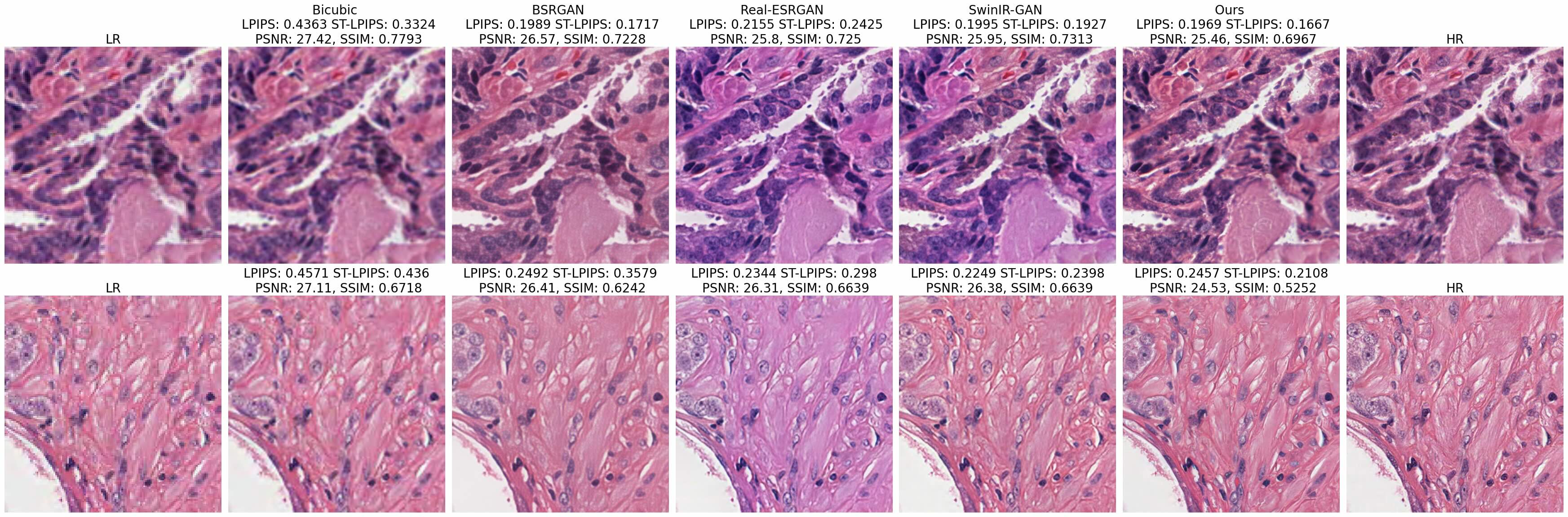}
\caption{Evaluation examples using PSNR, SSIM, and LPIPS in the field of digital pathology image super-resolution. Generative super-resolution methods produce images with sharper details and closer resemblance to high-resolution ground truths compared to bicubic interpolation. However, bicubic images often achieve higher PSNR, SSIM scores. Despite higher scores, bicubic images appear blurrier to human observers, indicating a disconnect between these metrics and human perception of image quality.}
\label{psnr}
\end{figure*}
% \begin{figure*}[!t]
% \centering
% \includegraphics[scale=.8]{figs/teaser_0502.png}
% \caption{Comparisons of state-of-the-art methods and our Histo-Diffusion.}
% \label{teaser0}
% \end{figure*}

% \begin{figure*}[!t]
% \centering
% \includegraphics[scale=.55]{figs/overview.png}
% \caption{Workflow.}
% \label{workflow}
% \end{figure*}
However, the technical limitations of scanning equipment and slide preparation variability make obtaining the high-resolution images a great challenge~\cite{farahani2015whole,zarella2019practical}. The resolution of WSIs is directly tied to the capabilities of the scanning equipment and some scanners may not have sufficiently high-resolution optics or sensors, especially the older or more budget-conscious models. Also, sometimes a compromise is made by resorting to lower resolution to keep file size manageable for image processing and transmission. Additionally, the quality of the initial slide preparation, including tissue sampling, processing and staining can greatly affect the final image resolution. Inconsistencies or deficiencies in any of above steps can lead to lower-quality images~\cite{smith2021developing,dunn2024quantitative}. Besides, the type and quality of staining techniques can influence the resolution and clarity of the images~\cite{runz2021normalization}. Uneven or poor staining can mask or blur important details, effectively reducing the usable resolution of the images.
Therefore, it is important to find a way to synthesize the high-resolution images from low-resolution ones while keeping the finer details and subtleties of tissue and cellular structures. Super-resolution (SR) offers a promising approach to addressing this challenge. Image super-resolution encompasses a suite of image processing techniques aimed at reconstructing high-resolution (HR) images from their low-resolution (LR) counterparts. 

Deep learning has significantly propelled advancements in image super-resolution within the natural image domain. Since 2017, Generative Adversarial Networks have been increasingly applied to image super-resolution tasks~\cite{ledig2017photo, goodfellow2020generative, ma2020pathsrgan, jose2021generative, manuel2022impact}. Their proficiency in synthesizing high-fidelity images has elevated them to prominence in this domain. Specifically for natural images, GANs can generate high-resolution images with remarkable detail, accurately simulating authentic high-resolution visuals. The adversarial training approach enables GANs to iteratively refine their generative capabilities, continually improving image quality to challenge the discriminator's ability to distinguish them as synthetic.

% However, training GANs is fraught with complexity, often leading to mode collapse, where the generator produces only a limited range of outputs, and non-convergence, where the generator and discriminator fail to reach equilibrium. Their performance depends heavily on the quality and diversity of the training data, making GANs particularly prone to overfitting and mode collapse when trained in the field of histopathology. In attempting to enhance image resolution, GANs can inadvertently introduce artifacts or excessively emphasize certain features, potentially causing misinterpretations that may lead to misdiagnosis if such exaggerated features are mistakenly identified as pathological anomalies~\cite{thanh2020catastrophic, saxena2021generative}.

However, training GANs often faces instability due to the adversarial interactions between the generator and discriminator. Such instability may lead to mode collapse, where the generator produces only a limited variety of outputs, and non-convergence, where the models fail to reach equilibrium~\cite{thanh2020catastrophic, saxena2021generative}. Consequently, GANs struggle to fully capture the diversity of histopathology images and to provide the finer details necessary for accurate diagnosis in histopathology. This presents significant challenges for realistic image synthesis and evaluation. 

Diffusion models have recently gained great success in natural image synthesis tasks ~\cite{dhariwal2021diffusion,ho2020denoising,nichol2021improved}. They are known for their ability to generate high-quality, realistic images. Intrinsically, diffusion models gradually denoise an image, starting from a random noise distribution. This process makes them inherently robust to noise, rendering them particularly efficacious in the enhancement of low-resolution or qualitatively compromised imagery. Their formidable generative capacities facilitate the interpolation of absent details through the synthesis of plausible textures and patterns, even when such elements are conspicuously unclear in the low-resolution images. In the context of digital pathology, this means producing super-resolved images that maintain the integrity and authenticity of the original biological structures, which is crucial for accurate diagnosis and analysis. Pathology images often contain complex textures and patterns that are essential for disease diagnosis. Diffusion models are particularly adept at handling these complexities, ensuring that the finer details of cellular structures are accurately represented in the super-resolution images. These models are inherently robust to noise and variations in the input data, an important feature when dealing with pathology images that may have inconsistencies due to different preparation techniques or imaging conditions. The robustness of diffusion models has significantly increased their utilization in digital pathology, particularly for applications such as data augmentation, synthetic data generation, and out-of-distribution detection~\cite{pozzi2023generating,linmans2024diffusion,oh2023diffmix}. By addressing challenges related to data scarcity and variability in histopathological images, these models enhance anomaly detection and ensure robustness in diagnostic systems through their ability to capture intricate patterns within pathology images. Although the potential of super-resolution techniques in the digital pathology field remains underexplored, the demonstrated capabilities of diffusion models suggest they can effectively capture complex structures and generate high-quality images in histopathology.

In the context of histopathology images, the accuracy and reliability of generated images are critical for downstream diagnostic and prognostic tasks. Therefore, robust evaluation metrics are essential to ensure the clinical viability of super-resolution images. Recently, tailored evaluation metrics have been proposed for medical image analysis, particularly in radiology~\cite{ssim_m,maruyama2023properties}. However, there is a lack of metrics specifically designed for histopathology image super-resolution. As a result, research in this area often relies on natural image quality assessment (IQA) metrics to evaluate histopathology super-resolution images~\cite{survey}. Metrics like PSNR (Peak Signal-to-Noise Ratio), SSIM (Structural Similarity Index Measure), and LPIPS (Learned Perceptual Image Patch Similarity)~\cite{lpips}, despite their widespread use in natural image quality assessments, can produce misleading results in histopathology, as illustrated in the caption details of Fig. \ref{psnr}.

Therefore, there is a pressing need for a methodology that works for histopathology image super-resolution while also providing reliable evaluation of the generated images. The super-resolution model should consider the intricate characteristics of digital pathology images and the histopathological microenvironment. Evaluation metrics should also accurately assess the similarity between generated super-resolution images and their high-resolution counterparts while effectively accommodating the variability in color, texture, and structure that results from diverse staining techniques and tissue types.

% Thus, a specialized evaluation metric for digital pathology super-resolution is urgently required to assess the similarity between generated super-resolution images and their high-resolution counterparts. This metric must accommodate variability in color, texture, and structure due to staining techniques and tissue types while being sensitive to the subtle morphological patterns critical for diagnostic accuracy.

In this paper, we first introduce Histo-Diffusion to generate super-resolution images specifically for digital pathology as shown in Fig.~\ref{teaser0}. Next, we present an effective methodology for evaluating the quality of these generated images.

Our key contributions are:

\begin{itemize}
    \item \textbf{Novel Evaluation Methodology:} We are the first to identify the sub-optimality in applying IQA metrics to the histopathology image super-resolution task. We point out the limitations of both full-reference and no-reference IQA metrics in evaluating histopathology images. These metrics fall short in assessing the complex microenvironment and structural details in histopathology images, which are crucial for accurate diagnosis. To address these issues, we propose a comprehensive evaluation methodology specifically tailored for digital pathology images. This methodology incorporates both full-reference metrics (comparing the generated super-resolution images to their ground truth counterparts) and no-reference metrics (trained on our histopathology IQA dataset). It evaluates fidelity, realism, and similarity to high-resolution ground truth images. Additionally, for the first time, we employ the CLIP model~\cite{clip,clipiqa} to assess image quality by measuring the alignment of generated images with human-like perception in histopathology.
    \item \textbf{Histopathology IQA Dataset:} We curated a histopathology image quality assessment dataset using the TCGA database, addressing the absence of available datasets for no-reference IQA metric training. Each image is assigned a quality score based on noise level, facilitating future histopathology IQA tasks and enabling accurate assessment of sharpness and noise in histopathology image super-resolution.
    \item \textbf{Histo-Diffusion:} We have proposed Histo-Diffusion, adapted from DiffBIR~\cite{diffbir}, as the first application of diffusion models for super-resolution image generation in digital pathology. Capable of handling multiple super-resolution scales and adaptable across various cancer types, Histo-Diffusion represents a significant advancement over current state-of-the-art methods.  
    \item \textbf{Detailed Comparative Analysis:} Our comprehensive comparative analysis highlights the effectiveness of Histo-Diffusion and provides an in-depth evaluation of diffusion versus GAN-based super-resolution methods.
    Across various cancer domains, Histo-Diffusion achieves the best ST-LPIPS scores, with improvements of 12.93\% for PRAD, 20.83\% for LUAD, and 12.88\% for GBM compared to the second-best GAN-based methods. Additionally, we enhance no-reference performance with MUSIQ score improvements of 13.26\% for PRAD, 16.40\% for LUAD, and 3.97\% for GBM. We also increase texture and intensity similarity by 17.96\% (PRAD-texture), 17.19\% (GBM-texture), 19.74\% (PRAD-intensity), and 9.77\% (GBM-intensity). This analysis provides critical insights for researchers, enabling them to identify the most suitable approach for their specific requirements.  
    %Across various cancer domains, Histo-Diffusion achieves best ST-LPIPS scores: 0.2079 (PRAD), 0.2250 (LUAD), and 0.3663 (GBM) compared to GAN-based methods. Additionally, we enhance no-reference performance, increasing MUSIQ scores to 45.00 (PRAD), 43.22 (LUAD), and 49.13 (GBM). This analysis delivers critical insights for researchers, enabling them to identify the most appropriate approach for their specific requirements.
\end{itemize}

\section{Related Work}
% In this section, we briefly discuss the histopathology image super resolution and nuclear segmentation.
\paragraph{\textbf{Histopathology Image Super-Resolution}}
Deep learning has become the predominant approach for super-resolution tasks in histopathology imaging. The specific challenges of super-resolution in this field began to receive notable attention around 2018, as evidenced by Mukherjee et al.'s use of CNNs for reconstructing high-resolution images in digital pathology~\cite{mukherjee2018convolutional}. This approach was expanded with the development of a recurrent CNN model designed to generate super-resolution images from multi-resolution WSI datasets~\cite{mukherjee2019super}. 
\begin{figure*}[!t]
\centering
\includegraphics[scale=.55]{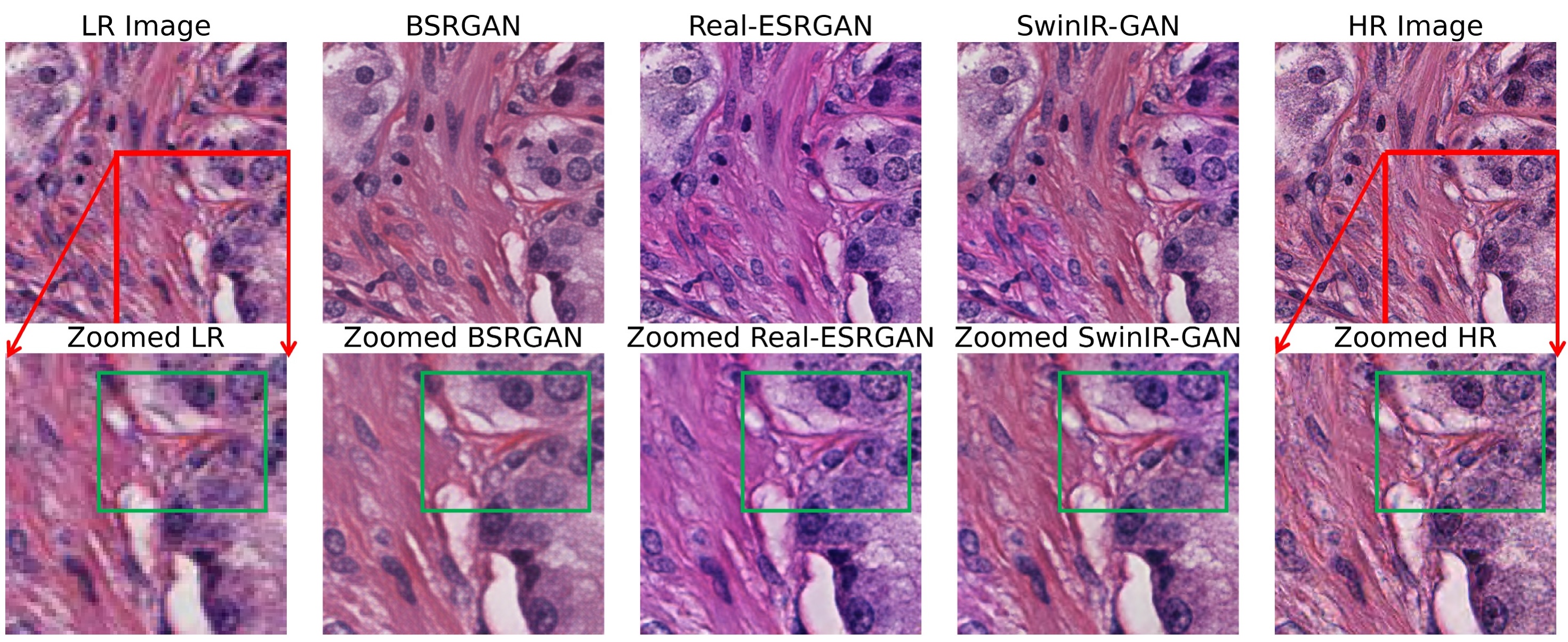}
\caption{Generated super-resolution images using GAN-based methods. These methods struggle to preserve stain color in histopathology images. The zoomed-in regions in the right corner of the high-resolution ground truth image, highlights that the GAN-generated super-resolution images exhibit over-smoothing and lack critical texture information within the green box, which are very crucial to accurate diagnosis. }
\label{related}
\end{figure*}

Concurrently, GANs were also being applied to image super-resolution tasks, with SRGAN~\cite{SRGAN} becoming the first framework to produce photo-realistic natural images for 4x upscaling factors using a perceptual loss function that combines adversarial and content loss. Subsequently, Enhanced SRGAN (ESRGAN)~\cite{esrgan}improved upon SRGAN by incorporating the Residual-in-Residual Dense Block (RRDB) without batch normalization. Further advancements were made with the introduction of Real-ESRGAN\cite{realesrgan}, which employed a sophisticated degradation modeling process to better simulate real-world image degradation. This technique has since become a popular method for super-resolution in histopathology imaging~\cite{rong2023enhanced}.

In 2020, the introduction of vision transformers significantly advanced performance in imaging tasks, leading to the creation of SwinIR~\cite{swinir}, a model specifically designed for image restoration. SwinIR has demonstrated notable success in real-world super-resolution scenarios~\cite{puttagunta2022swinir,zhang2022swinfir,choi2023n}. 

~\cite{survey} conducted a comparative study in histopathology image super-resolution, using CNN-based and GAN-based models to assess the quality of super-resolved histopathology images. 
% It also concluded that the Image Quality Assessment (IQA) metrics commonly used in natural image analysis do not effectively translate to histopathology images, emphasizing the need for specialized metrics in this domain.
As depicted in Fig ~\ref{related}, GAN-based methods struggle to preserve stain color and intricate microenvironment details when generating super-resolution histopathology images.

% Consequently, a lack of research persists in addressing the intricate and complex microenvironment characteristics of histopathology images in super-resolution. This gap presents a promising area for future exploration that targets the unique challenges specific to histopathology image super-resolution.

\paragraph{\textbf{Diffusion Models}}
Since their inception around 2020, diffusion models have become increasingly prominent across various fields of image generation and restoration, due to their ability to produce high-quality, coherent images. A pivotal development in this domain is the Denoising Diffusion Probabilistic Models (DDPMs)~\cite{ho2020denoising}. DDPM employs a Markov chain to incrementally convert noise into samples from the data distribution, substantially boosting the generative prowess of diffusion models and notably enhancing the quality of the output. The Score-Based Generative Model~\cite{song}, innovates by utilizing stochastic differential equations to model the data distribution's gradient. This model has demonstrated its competitive edge, proving capable of generating high-quality samples across diverse applications.

SR3~\cite{sr3}, the first super-resolution model based on diffusion models,  iteratively upgrades low-resolution images to high-resolution outputs. It excels at managing large upscaling factors, such as 8x and 16x, while maintaining high image fidelity. Cascaded diffusion models~\cite{cascaded}, from the same group, consist of a series of diffusion models each designed to stepwise enhance the image resolution. This strategy allows for more controlled image refinement, thereby improving detail retention in super-resolved images. 

Concurrently, the Latent Diffusion Model (LDM)\cite{ldm} was proposed, quickly becoming the standard for high-resolution image synthesis. Unlike SR3 and cascaded models, LDM operates within a compressed latent space, enabling diffusion models to scale efficiently and produce high-resolution images with less computational overhead while still delivering remarkable image quality. On this foundation, the same team~\cite{ldm} developed Stable Diffusion, which merges latent diffusion models with techniques from variational autoencoders (VAEs) conditioned on textual prompts. This process facilitates the creation of images that closely mirror the descriptions, establishing it as the current SOTA method for generating high-quality images from textual descriptions. This advancement has also provide a way for applying diffusion models in the field of histopathology imaging.~\cite{xu2023vit,pathldm,srikar2}.

Expanding on this framework, ControlNet~\cite{controlnet} was developed to embed additional control mechanisms beyond textual descriptions into stable diffusion. It leverages the generative diffusion prior of stable diffusion, trained on natural images, and demonstrates effective control over stable diffusion in image generation with various conditioning inputs such as canny edges, user scribbles, and human poses. This control mechanism makes ControlNet highly effective in natural image super-resolution, and a few studies~\cite{diffbir,stablesr} have begun incorporating low-resolution images as control inputs to generate high-resolution images. They utilize stable diffusion's powerful generative prior with an architecture similar to ControlNet. Because of its robustness against noise and its ability to produce high-fidelity super-resolution images, these approaches has found success in natural image super-resolution.

Consequently, ControlNet offers a control mechanism for the generation process in image super-resolution without the need to retrain the stable diffusion model, which demands significant time and computational resources. It enables the application of stable diffusion to super-resolution tasks. In the context of histopathology image super-resolution, the key challenge is to harness the diffusion prior and include histopathological information to generate high-resolution images that preserve essential contextual information and complex microenvironments, critical for accurate diagnosis.

\paragraph{\textbf{Evaluation Metrics in Super-Resolution}} Image quality assessment metrics are divided into two categories: full-reference IQA metrics, which compare a generated image to a high-quality reference image, and no-reference IQA metrics, which evaluate image quality without any reference.

Commonly used full-reference metrics for natural images include PSNR, SSIM, and LPIPS. PSNR measures the discrepancy between a super-resolution image and its high-resolution counterpart by calculating the peak signal-to-noise ratio. While useful for detecting overall pixel errors, PSNR is overly sensitive to noise, which is prevalent in pathological images due to staining and scanning artifacts. Simple pixel-wise comparisons fail to capture the nuanced tissue structures and cellular morphology necessary for diagnosis. SSIM examines changes in structural information, luminance, and contrast to assess image quality. Although better aligned with human perception than PSNR, SSIM might overlook crucial histopathological features like complex micro-environment and structural information, which are essential for accurate diagnosis. SSIM also struggles to evaluate images at different scales or resolutions, which is a frequent challenge in digital pathology. LPIPS~\cite{lpips}, a recent deep-learning metric used to evaluate perceptual similarity, depends heavily on the characteristics of its training data. ShiftTolerant-LPIPS (ST-LPIPS)~\cite{shift}, developed on the foundation of the LPIPS metric, is an enhanced perceptual similarity metric that integrates tolerance to small spatial shifts, thereby increasing its robustness and reliability for various image comparison tasks. Because these models were primarily trained on natural images, they may not fully comprehend the nuances of pathology images and could miss critical structures. Figure \ref{psnr} illustrates the limitations of PSNR, SSIM, and LPIPS in digital pathology super-resolution. Images produced via bicubic interpolation can, interestingly, score higher in these metrics despite being blurry, emphasizing the need for metrics that can accurately evaluate texture and intensity, which are crucial for clinical interpretation. 

Another possible approach for evaluating generated super-resolution images is no-reference metrics. While full-reference IQA metrics evaluate the resemblance between original high-resolution images and generated super-resolution images, no-reference IQA (also known as blind IQA) assesses images without a reference. It focuses on inherent features, making it especially useful in scenarios where no ideal reference image is available. However, no-reference IQA requires datasets with image-quality score pairs for training, which are not available in histopathology.

\section{Methodology}
\begin{figure*}[!t]
\centering
\includegraphics[scale=.55]{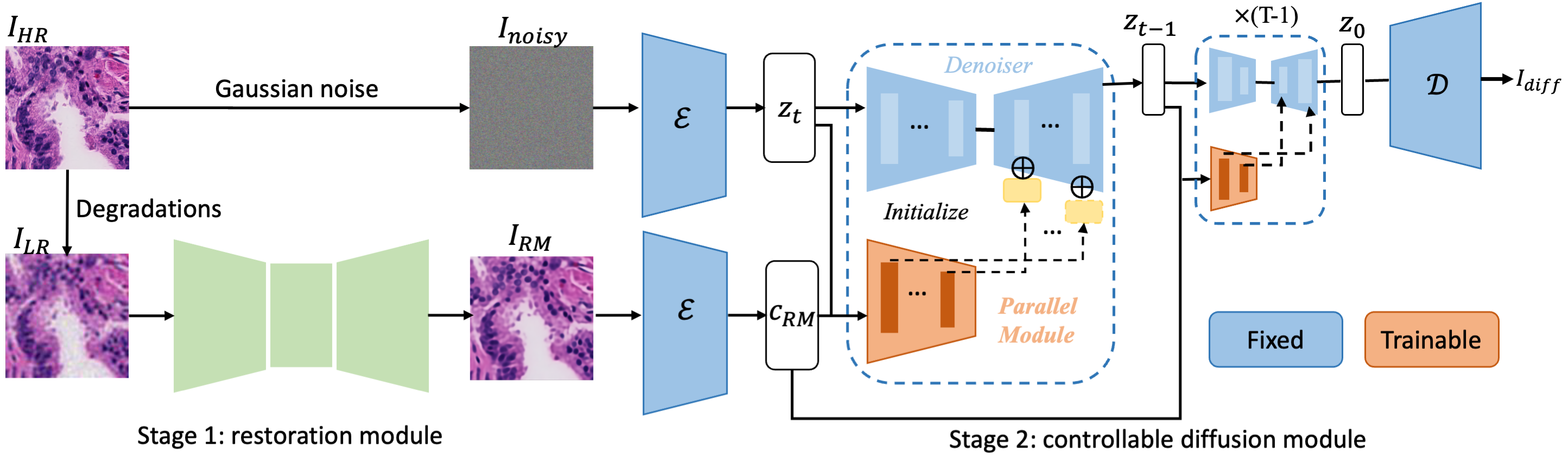}
\caption{Dual-Stage Diffusion-Based Image Super-Resolution Model. It includes a restoration module that generates restored images as histopathology priors for a controllable diffusion module. The restored and noisy latent images are combined to work as the input for the controllable diffusion module for super-resolution image generation. }
\label{workflow}
\end{figure*}
The task of image super-resolution (SR) can be formally defined as a process of generating a high resolution (HR) image from low-resolution (LR) observations of the same scene. The goal is to reconstruct a high-fidelity image that is as close as possible to the original, unseen high-resolution image, both in terms of pixel values and perceptual quality.

\textbf{Given:}
\begin{itemize}
    \item An input low-resolution image $I_{LR}$, which is typically a downsampled version of a high-resolution image $I_{HR}$, possibly also degraded by factors such as blur, noise and compression artifacts.
    \item A desired upscaling factor $s$, which specifies how much larger the high-resolution image should be compared to the low-resolution input. This factor is usually expressed as a multiplier for both the width and height dimensions ($s\times width, s\times height)$.
\end{itemize}

\textbf{Target:}
\begin{itemize}
    \item To construct a super-resolution image $I_{SR}$ that maximizes the fidelity to the original high-resolution source image $I_{HR}$, from which $I_{LR}$ was derived. 
    
\end{itemize}

Drawing inspiration from~\cite{diffbir}, we adopt a dual-stage framework for histopathology image super-resolution task. This framework includes a restoration module to provide histopathology prior and a controllable diffusion module derived from ControlNet~\cite{controlnet} for histopathology image generation. This dual-stage approach addresses the super-resolution challenges specific to histopathology images. The restoration module reduces the degradations and generates a histopathology-specific prior, serving as the condition for the controllable diffusion module. This controllable diffusion module ensures stable diffusion by utilizing models pretrained on natural images, while the restoration module customizes the stable diffusion specifically for histopathology applications. We also propose
a comprehensive evaluation strategy to tackle the difficulties associated with applying traditional image quality assessment metrics to the histopathology imaging. Overall, our paper offers a holistic solution for the generation and evaluation of super-resolution histopathology images.

\subsection{Restoration module}

To accurately replicate the complex microenvironments and artificial noise characteristics encountered in histopathology image super-resolution, we employ a restoration module utilizing the SwinIR~\cite{swinir} model to refine the fidelity and details of these degraded images.

We simulate the degradation process on high-resolution images, $I_{HR}$, employing advanced degradation techniques. Histopathology image super-resolution challenges include the introduction of noise, artifacts, and the complexities of the microenvironment. Key degradation techniques such as blurring, resizing, and the introduction of noise are utilized to produce corresponding low-resolution image, $I_{LR}$. The SwinIR model is then applied to reduce the effects of these degradations, particularly focusing on noise and compression artifacts.

The image restoration process includes three principal stages: shallow feature extraction, deep feature extraction, and high-quality image reconstruction. We follow the modified SwinIR~\cite{diffbir} approach by first downsampling the original low-resolution input image using a pixel unshuffle operation with a scale factor of 8, followed by a convolutional layer for shallow feature extraction. For deep feature extraction, we employ Residual Swin Transformer Blocks (RSTB), each containing several Swin Transformer Layers (STL). To upsample the deep features, nearest neighbor interpolation is performed three times, with each step followed by a convolutional layer and a leaky ReLU activation layer. This process restores the image $I_{LR}$ to its original dimensions, resulting in the final restored image $I_{RM}$.

The restoration module is optimized by minimizing 
MSE  loss as defined below:

\begin{equation}
I_{RM} = \mathrm{RM}(I_{LR}), \quad \mathcal{L}_{RM} = ||I_{RM} - I_{HR}||^2_2
\label{swinir}
\end{equation}

Here, $I_{RM}$ is derived through regression learning and subsequently employed to enhance the performance of the controllable diffusion module. This restoration module not only mitigates the impact of initial image degradations but also facilitates the generation of super-resolution images that are closer in quality to the original high-resolution counterparts.

\begin{figure}[!t]
\centering
\includegraphics[scale=.4]{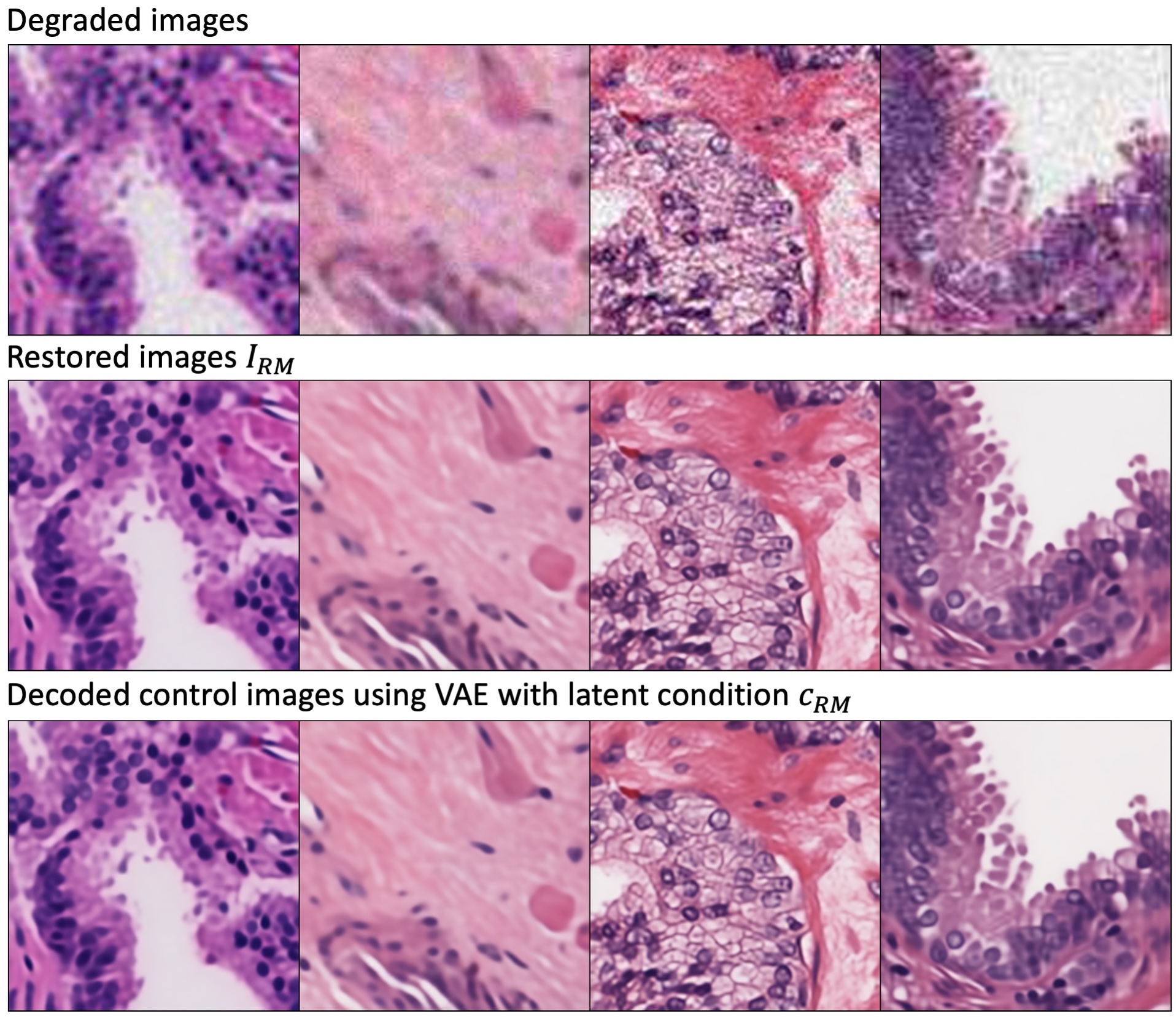}
\caption{Restored images with corresponding decoded control images using VAE with condition latent $c_{RM}$}
\label{decodecontrol}
\end{figure}

\begin{figure}[!t]
\centering
\includegraphics[scale=.25]{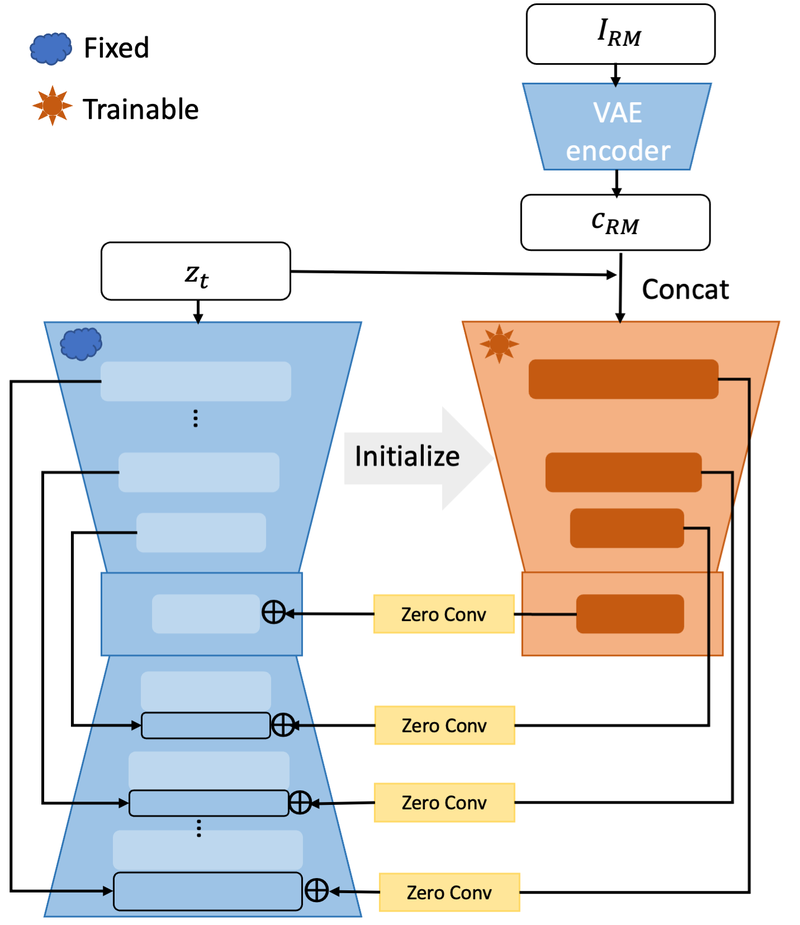}
\caption{Architecture of the controllable diffusion module. The right orange section represents the trainable ControlNet, while the left blue section indicates the fixed UNet.  The condition latent $c_{RM}$ is combined with noisy latent $z_t$ to control the generation process of super-resolution images.}
\label{control}
\end{figure}

\begin{figure*}[!t]
\centering
\includegraphics[scale=.5]{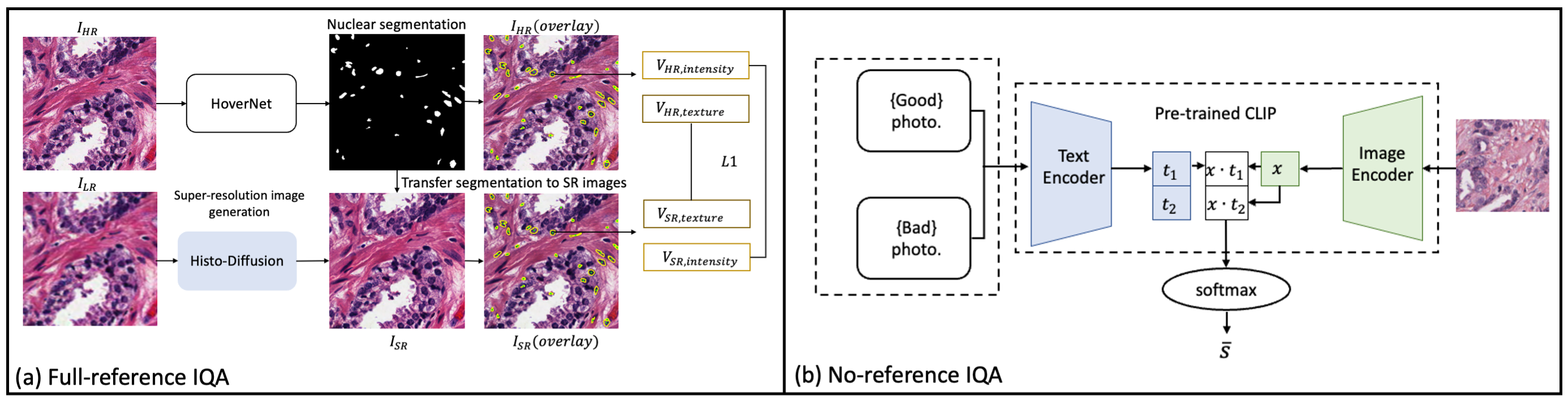}
\caption{Proposed IQA metric pipeline. (a) Full-reference IQA metric: High-resolution ground truth images are processed through HoverNet to obtain nuclear segmentation. Within these nuclear positions, texture and intensity information of these ground truth images are then compared to those of the generated super-resolution images.(b) No-reference IQA metric: We leverage CLIP-IQA model to assess image quality with our own curated histopathology IQA dataset.}
\label{nuclear}
\end{figure*}

\begin{figure*}[!t]
\centering
\includegraphics[scale=.5]{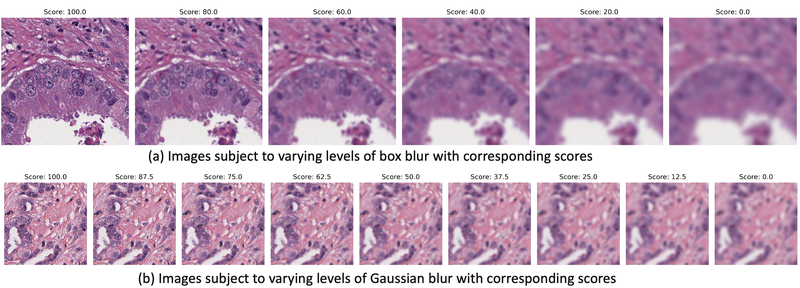}
\caption{Images subject to varying blur levels with corresponding Scores}
\label{blur}
\end{figure*}

\subsection{Controllable diffusion module}

After obtaining the restored image $I_{RM}$, we use it as the conditioning input for our controllable diffusion module. Leveraging the pre-trained Stable Diffusion model, the restored image serves as the condition to generate super-resolution histopathology images, as shown in Fig. ~\ref{control}. Our controllable diffusion module (CDM) involves two key steps: 1) encoding the restoration module into condition latent; and 2) using this condition latent to control the generation of super-resolution histopathology images.

In the first step, we take the restored image $I_{RM}$ and leverage the VAE from Stable Diffusion to encode it into a latent space~\cite{vqgan}, denoted as $c_{RM} = \mathcal{E}(I_{RM})$. As illustrated in Fig.~\ref{decodecontrol}, the decoded images closely resemble the restored images, demonstrating that the VAE, trained on a large-scale dataset, can accurately reconstruct the restored image $I_{RM}$. The conditioned latent $c_{RM}$ captures sufficient image information from the restoration module, preserving all critical details for use in the subsequent super-resolution image generation process.

In the second step, we use the conditioned latent $c_{RM}$ to control the Stable Diffusion generation process. Following~\cite{controlnet,diffbir}, we make a trainable copy of the pre-trained UNet encoder and middle block, referred to as $\mathbf{F}_{cond}$. The conditioned latent $c_{RM}$ is concatenated with the noisy latent $z_t$ at time $t$ to form the combined latent representation $z_{t}^{'} = \text{cat}(z_t, c_{RM})$, where $\text{cat}$ denotes the concatenation operation. The combined latent $z_{t}^{'}$ is fed into the control net (orange part as shown in Fig.~\ref{control}). In the default configuration of stable diffusion, the UNet in the Stable Diffusion blue region only accepted the noisy latent $z_t$, However, with the introduction of ControlNet in the orange region, it now accepts the combined latent $z_{t}^{'}$ as the condition. The first layer's channel number has been modified to accommodate the combined latent dimension. Feature modulation occurs solely at the middle block and through skipped connections, utilizing addition operations. Zero convolutions are strategically employed to bridge the connection between the yellow ControlNet and the fixed UNet denoiser.

During training, $c_{RM}$ serves as the histopathology prior, and the text prompt $c$ is set to blank (``''). Only the parameters of the control net and the feature modulation are optimized by minimizing the latent diffusion objective, as shown in equation~\ref{cdm}. This process refines the quality and resolution of the generated images based on the structured guidance provided by the concatenated latent inputs.

This targeted approach ensures that the network $\epsilon_\theta$ is trained to predict the noise $\epsilon$ using $c_{RM}$ and a blank text prompt $c$, effectively learning to enhance image resolution while maintaining the integrity and details necessary for accurate histopathological analysis.

\begin{equation}
    \mathcal{L}_{CDM}=\mathbb{E}_{z_t,c,t,\epsilon,c_{RM}}[||\epsilon-\epsilon_\theta(z_t,c,t,c_{RM})||^2_2].
\label{cdm}
\end{equation}

\subsection{Evaluation strategy}
Current image quality assessment metrics often fall short when applied to super-resolution tasks in histopathology imaging. This is because these metrics typically rely on large natural image datasets (no-reference IQA) and fail to accurately evaluate the texture and intensity, which are crucial for clinical interpretation. In response to this deficiency, we propose a comprehensive evaluation strategy tailored specifically for digital pathology images. This strategy encompasses both full-reference and no-reference image quality assessments, designed to accurately gauge the performance of super-resolution techniques in a context where precise detail and image fidelity are paramount. This dual approach allows for a more holistic assessment of image quality, addressing both the comparison of super-resolved images to high-resolution ground truths and the intrinsic qualities of images when ground truths are unavailable.

\paragraph{\textbf{Full-reference image quality assessment}}
In the realm of full-reference image quality assessment within digital pathology, our paper focuses on evaluating super-resolution images $I_{SR}$, against their corresponding high-resolution ground truth images, $I_{HR}$. Given the critical role of nuclei segmentation in the application of super-resolution techniques to histopathology images, it is imperative to determine whether $I_{SR}$ can accurately replicate the texture and intensity characteristics of $I_{HR}$. For nucleai segmentation, we employed HoVer-Net~\cite{hover}, a commonly utilized nuclear segmentation model, pretrained on the CoNSeP dataset which includes 41 hematoxylin and eosin (H\&E) stained image tiles, each $1,000\times1,000$ pixels, captured at a 40x objective magnification. The process involves inputting $I_{HR}$ into HoverNet to identify nuclear locations, which are subsequently used to analyze corresponding areas in $I_{SR}$, as illustrated in Fig~\ref{nuclear} (a).

For each generated $I_{SR}$, we conduct a detailed comparison of nuclei position, intensity, and texture properties against $I_{HR}$. For each nucleus, we calculate mean, standard deviation, skewness, and kurtosis of grayscale intensity values to form the intensity feature vector $V_{SR,intensity}$. Similarly, for texture, we measure contrast, dissimilarity, homogeneity, and energy, comprising the texture feature vector $V_{SR,texture}$. These vectors are compared to their respective ground truth vectors $V_{HR,intensity}$ and $V_{HR,texture}$ using the $L1$ metric, yielding $L1_{intensity}$ and $L1_{texture}$ differences, respectively. We employ these $L1$ values as full-reference metrics to quantify the similarity between the generated super-resolution images and the ground truth high-resolution images.

This full-reference $L1$ metric allows us to quantitatively evaluate the fidelity of $I_{SR}$ images in replicating critical nuclear details, assessing their similarity in texture and intensity to $I_{HR}$. This comparison enables us to determine which super-resolution method most accurately reflects the nuanced nuclear properties observed in high-resolution images, thus validating our super-resolution techniques both theoretically and in practical, clinical settings.

\paragraph{\textbf{No-reference image quality assessment}}

Current no-reference image quality assessment (IQA) metrics are predominantly tailored for evaluating natural images by analyzing how closely an input image resembles real-world imagery. However, these metrics often fail to capture the complex textures and structures characteristic of histopathology images. To bridge this gap, we have developed a specialized histopathology image dataset with quality scores derived from the TCGA-PRAD database. We selected 5000 patches at $40\times$ magnification, representing the high-resolution ground truth with a maximum score of 10.0. For each image in this set, we introduced varying levels of noise using two types of distortion (box blur and Gaussian blur). For the box blur, the noise level was incrementally increased by adjusting the radius, with scores assigned based on the corresponding radius level. Similarly, for the Gaussian blur, the standard deviation of the Gaussian kernel was gradually increased, resulting in deteriorating scores. We curated two datasets of histopathology patches—one for box blur and the other for Gaussian blur—each rated on a scale from 10.0 (best) to 0.0 (worst, most noisy). The scores are illustrated in Fig~\ref{blur}. 

The rationale behind selecting these specific types of noise is that generative models in histopathology image super-resolution often suffer from over-smoothing, resulting in super-resolved images that lack sharpness and detail. By employing box blur and Gaussian blur, we aim to evaluate the extent of blur and over-smoothing in generated super-resolution images. Consequently, we have created two distinct histopathology image datasets with quality scores for subsequent training of the IQA model.

To effectively learn and predict these quality scores for histopathology images, we utilize the state-of-the-art CLIP-IQA model~\cite{clipiqa}. CLIP-IQA leverages the capabilities of the CLIP model to assess image quality through contrastive assessment and semantic alignment. As illustrated in Fig~\ref{nuclear} (b), CLIP-IQA uses images with quality scores as inputs. Each image is processed by the CLIP model to extract feature embeddings. Similarly, textual descriptions that represent high-quality reference standards are also converted into embeddings using the CLIP model. The quality of an image is determined by comparing its embedding with those of the reference standards, which represent various quality attributes or common defects such as sharpness, color fidelity, and blurring. Higher similarity scores indicate better alignment with the reference standards, suggesting higher perceptual quality.

We trained CLIP-IQA with our specifically curated dataset, enabling it to recognize and quantify the degradation in histopathology images due to blurring effects. After training, the CLIP-IQA model is employed to assess the quality of images during testing, providing a robust mechanism to evaluate image quality in a domain-specific context. The generated super-resolution images are then input into the trained CLIP-IQA to obtain quality scores CLIP-IQA (boxblur), CLIP-IQA (Gaussian), ensuring that the assessment is not only precise but also highly relevant to the specific requirements of histopathology image analysis.

\begin{figure*}[!t]
\centering
\includegraphics[scale=.5]{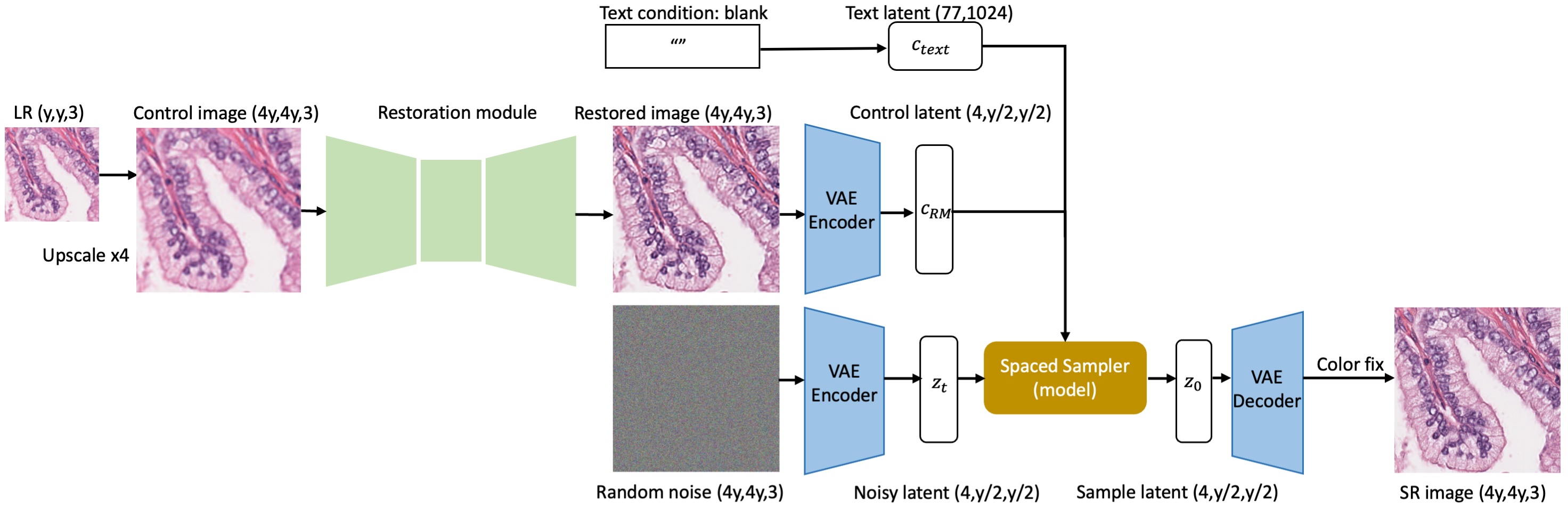}
\caption{Inference process of Histo-Diffusion. During inference, a low-resolution image with dimensions $(y, y, 3)$ is fed into the model for image super-resolution generation. Due to the capability of the stable diffusion's UNet to handle any latent whose dimensions  are multiples of 8, our diffusion-based super-resolution model can process any size input images. The low-resolution image is resized to the target upscaled size $(4y,4y,3)$  and preprocessed using the SwinIR-based restoration module to generate a restored image. This restored image is downsampled via the VAE encoder to produce the control latent. The control latent, with dimensions $(4, y/2, y/2)$, must be a multiple of 8 for the UNet in Stable diffusion, which means the control image $(4y, 4y, 3)$ must be a multiple of 64. If the control image dimensions are not multiples of 64, padding is applied to the control image to meet this requirement.
Concurrently, a random noisy image with the same dimensions as the restored image is generated and encoded by the VAE to obtain the noisy latent. A blank text condition (``'') is input into CLIP to derive the text embedding. Both the control latent and text embedding are fed into the trained model with a spaced sampler to generate sample latent, which are subsequently decoded by the decoder. Color fix is then applied to these samples to finally produce the super-resolution image.}

\label{infer}
\end{figure*}

\section{Experiments and Results}

\subsection{Datasets and Evaluation Metrics}

\paragraph{\textbf{Datasets}} We utilize the TCGA-PRAD dataset for curating our high-resolution (HR) and low-resolution (LR) patch-level dataset. A total of 200 WSIs were randomly selected, with an additional 70 WSIs forming our test set. No WSIs overlap between the training and test sets. The WSI ids are provided in the supplementary materials. These WSIs were tilled into smaller patches, yielding a training set of 200,000 patches and a validation set comprising 100 patches, all randomly selected from the initial 200 WSIs.
\begin{table}[t]
\small
\centering
\caption{Dataset distribution }
\resizebox{0.28\textwidth}{!}{%
\begin{tabular}{c| c c c }
\hline
 & Cancer type& \#Patches &\#WSIs\\\hline \hline
Train &PRAD& 200000 & 200\\\hline
Val &PRAD& 100 & 36\\\hline
\multirow{4}{*}{Test} & PRAD & 2000 & 70 \\
 & PRAD & 100 & 70 \\
 & LUAD & 100 & 6 \\
 & GBM & 100 & 17 \\\hline\hline

\end{tabular}}
\label{table_distribution}
\end{table}
For the purposes of evaluation, our dataset includes 100 TCGA-PRAD patches (PRAD-100), representing the cancer type used in training, and an expanded set of 2000 TCGA-PRAD patches (PRAD-2000) to encompass a broader testing cohort.  All patches were randomly selected from 70 WSIs in TCGA-PRAD. We also incorporate 100 TCGA-LUAD (LUAD-100) patches from 6 WSIs and 100 TCGA-GBM (GBM-100) patches from 17 WSIs to assess the model’s generalization capabilities across different cancer types. For LUAD-100, each patch consists of a HR image at $40\times$ magnification ($512\times512$ pixels) and a corresponding LR image at 10x magnification ($128\times128$ pixels). GBM-100 includes HR images at $20\times$ magnification ($512\times512$ pixels) and LR images at 5x magnification ($128\times128$ pixels) to further test the model's ability to handle different magnification scales. The distribution of the entire dataset is detailed in Table~\ref{table_distribution}.

\begin{figure*}[!t]
\centering
\includegraphics[scale=.73]{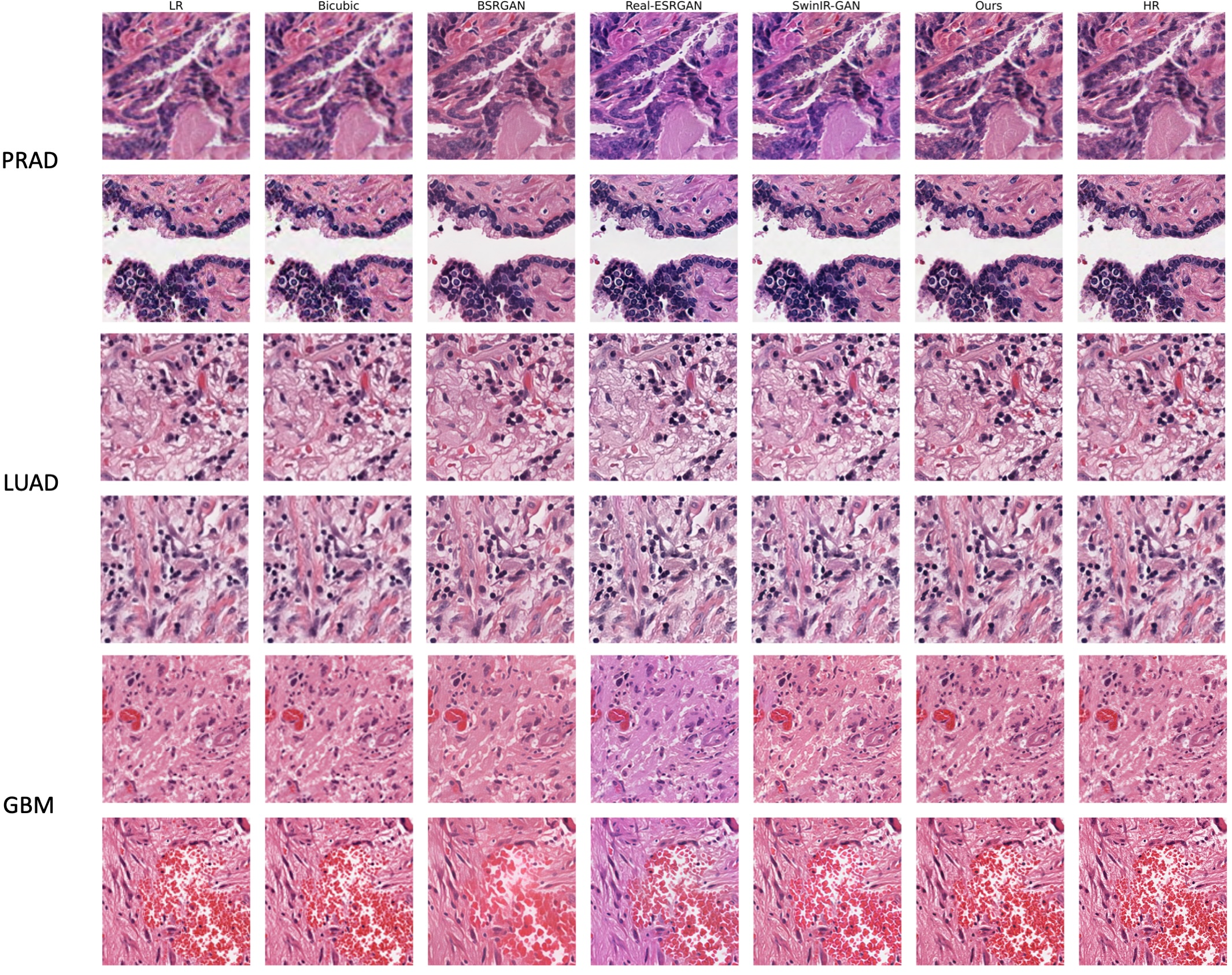}
\caption{ Visual comparisons on PRAD-100, LUAD-100 and GBM-100 samples. Please zoom in for more details.}
\label{teaser_0611}
\end{figure*}

\begin{table*}[t]
\small
\centering
\caption{Comparison with state-of-the-art GAN-based methods on histopathology image datasets with a 4× upsampling scale.  The best and second best results are highlighted in \textbf{bold} and \underline{underline}. }

\resizebox{0.98\textwidth}{!}{%
\begin{tabular}{c|c|c|c c c c c c c c c c c}
\hline
\multirow{2}{*}{Datasets}&\multirow{2}{*}{Methods} &\multirow{2}{*}{Scale} & \multicolumn{9}{c}{Metrics}\\\cline{4-13} 
& & & & PSNR$\uparrow$ & SSIM$\uparrow$ &LPIPS$\downarrow$& ST-LPIPS$\downarrow$ & CLIP-IQA$\uparrow$ & MUSIQ$\uparrow$ &NIQE$\downarrow$ & BRISQUE$\downarrow$ & NRQM$\uparrow$\\
\hline
\hline
\multirow{5}{*}{PRAD-100} &Bicubic& $\times4$ & &\textbf{26.48} & \textbf{0.6656} & 0.4534&0.4702&0.3738&25.96&8.64&64.09&3.18\\ 
&BSRGAN& $\times4$ & &\underline{25.78}&0.6335&0.2612&0.3462&0.3851&35.93&8.57&\underline{39.81}&4.38\\ 
&Real-ESRGAN& $\times4$ & &25.49&\underline{0.6639}&\underline{0.2277}&0.2598&\textbf{0.4753}&\underline{39.73}&8.82&46.71&4.61\\ 
&SwinIR-GAN& $\times4$ & &25.52&0.6605&\textbf{0.2229}&\underline{0.2388}&0.4499&38.87&\underline{8.56}&44.78&\underline{4.95}\\\cline{2-13}
&Ours& $\times4$ & &24.86&0.5947&0.2279&\textbf{0.2079}&\underline{0.4559}&\textbf{45.00}&\textbf{4.95}&\textbf{31.39}&\textbf{6.34}\\ 
\hline
\hline
\multirow{5}{*}{LUAD-100} &Bicubic& $\times4$ & & \textbf{28.27}&\textbf{0.7691}&0.3462&0.3923&0.2640&30.52 &7.62&56.41&3.44\\ 
&BSRGAN& $\times4$ & &26.85&0.7154&0.3307&0.3085&0.3585&36.04&7.36&\underline{44.29}&\underline{3.98}\\ 
&Real-ESRGAN& $\times4$ & & 25.76&0.7311&0.3300&0.3121&0.4135&\underline{37.13}&7.66&51.05&3.85\\ 
&SwinIR-GAN& $\times4$ & &\underline{27.09}&\underline{0.7577}&\underline{0.3113}&\underline{0.2842}&\underline{0.4177}&35.73&\underline{7.96}&48.49&3.88\\\cline{2-13}
&Ours& $\times4$ & &25.92&0.6666&\textbf{0.2526}&\textbf{0.2250}&\textbf{0.4309}&\textbf{43.22}&\textbf{5.12}&\textbf{33.14}&\textbf{5.15}\\ 
\hline
\hline
\multirow{5}{*}{GBM-100} &Bicubic& $\times4$ & & \textbf{24.34}&\textbf{0.5808}&0.5199&0.5484&0.4540&33.44&\underline{8.15}&52.83&3.05\\ 
&BSRGAN& $\times4$ & &\underline{23.39}&0.5199&0.3462&0.4790&0.3997&40.91&9.06&\underline{38.19}&4.22\\ 
&Real-ESRGAN& $\times4$ & & 24.08&0.5846&0.3634&0.4644&\textbf{0.5205}&44.51&9.34&39.95&4.22
\\ 
&SwinIR-GAN& $\times4$ & &24.31&\underline{0.6102}&\underline{0.3436}&\underline{0.4205}
&\underline{0.4954}&\underline{47.25}&10.05&40.53&\underline{4.42}\\\cline{2-13}
&Ours& $\times4$ & &22.88&0.4950&\textbf{0.2894}&\textbf{0.3663}&0.4862&\textbf{49.13}&\textbf{4.94}&\textbf{27.52}&\textbf{5.96}
\\ 
\hline
\hline
\multirow{5}{*}{PRAD-2000} &Bicubic& $\times4$ & &\textbf{27.05}&\textbf{0.6823}&0.4398& 0.4438 &0.3760&26.21 & 8.64 & 62.98 & 3.24
 \\ 
&BSRGAN& $\times4$ & &\underline{26.26}&0.6479&0.2529& 0.3268 &0.3886&35.76 & 8.52 & 37.99 & 4.43
  \\ 
&Real-ESRGAN& $\times4$ & &25.95&\underline{0.6759}&0.2235&0.2558 &\textbf{0.4805}&\underline{39.53} & 8.76 & 45.97 &4.64
\\ 
&SwinIR-GAN& $\times4$ & &26.05&0.6740&\textbf{0.2202}& \underline{0.2377} &0.4542&38.87&\underline{8.49} & \underline{44.97} & \underline{4.90}
\\\cline{2-13}
&Ours& $\times4$ & &25.36&0.6075&\underline{0.2221}&\textbf{0.2006}&\underline{0.4543}&\textbf{44.44} & \textbf{4.99} & \textbf{30.69} &\textbf{6.30}
\\ 
\hline
\hline
\end{tabular}}
\label{table_6metric}
\end{table*}

\begin{table*}[t]\small
\centering
\caption{ Comparison with state-of-the-art GAN-based methods on histopathology image datasets with a 4× upsampling scale on PRAD-100, LUAD-100, GBM-100, PRAD-2000 dataset using our proposed metrics.  The best and second best results are highlighted in \textbf{bold} and \underline{underline}. }
\label{table1}
\resizebox{0.7\textwidth}{!}{%
\begin{tabular}{c|c|c|c c c c c}
\hline
\multirow{2}{*}{Datasets} &\multirow{2}{*}{Methods} &\multirow{2}{*}{Scale} & \multicolumn{4}{c}{Metrics}\\ \cline{4-8}
 & & & & $L1_{texture}\downarrow$ & $L1_{intensity}\downarrow$ & CLIP-IQA (boxblur)$\uparrow$ & CLIP-IQA (Gaussian) $\uparrow$ \\ \hline
 \hline

\multirow{5}{*}{PRAD-100} & Bicubic& $\times4$ & &\underline{274.21}&\underline{14.74}&0.8109 &0.8197\\
 &BSRGAN& $\times4$ & &348.65&20.77&0.9015&0.8776\\ 
 &Real-ESRGAN& $\times4$ & &321.15&15.38&0.8884&0.8974\\ 
 &SwinIR-GAN& $\times4$ & &321.94&15.14&\underline{0.9304}&\underline{0.9288}\\\cline{2-8}
 &Ours& $\times4$ & &\textbf{224.95}&\textbf{11.83}&\textbf{0.9468}&\textbf{0.9458}\\\hline 
 \hline

\multirow{5}{*}{LUAD-100} & Bicubic& $\times4$ & &\textbf{160.41}&\textbf{11.04}&0.8768&0.8943 \\
 &BSRGAN& $\times4$ & &242.59&19.05&0.9010&0.8934\\ 
 &Real-ESRGAN& $\times4$ & & 271.38&16.59&\underline{0.9359}&\underline{0.9364}\\ 
 &SwinIR-GAN& $\times4$ & & 253.44&16.01&0.9307&0.9355\\\cline{2-8}
 &Ours& $\times4$ & &\underline{190.49}&\underline{13.14}&\textbf{0.9497}&\textbf{0.9504}\\\hline 
 \hline
 
\multirow{5}{*}{GBM-100} & Bicubic& $\times4$ & &\underline{388.57}&\underline{18.00}&0.7784 &0.7566\\
 &BSRGAN& $\times4$ & &470.60&29.81&0.9245&0.8965\\ 
 &Real-ESRGAN& $\times4$ & & 425.01&20.84&0.9032&0.9026\\ 
 &SwinIR-GAN& $\times4$ & & 397.22&19.38&\underline{0.9419}&\underline{0.9276}\\\cline{2-8}
 &Ours& $\times4$ & &\textbf{321.75}&\textbf{16.24}&\textbf{0.9513}&\textbf{0.9343}\\\hline 
 \hline

 \multirow{5}{*}{PRAD-2000} & Bicubic& $\times4$ & &\underline{257.04}&\underline{14.19}&0.8045 &0.8177\\
 &BSRGAN& $\times4$ & &331.23&20.27&0.9029&0.8791\\ 
 &Real-ESRGAN& $\times4$ & & 300.24&14.83&0.8908&0.8994\\ 
 &SwinIR-GAN& $\times4$ & & 304.11&14.66&\underline{0.9308}&\underline{0.9308}\\\cline{2-8}
 &Ours& $\times4$ & &\textbf{220.96}&\textbf{11.77}&\textbf{0.9464}&\textbf{0.9467}\\\hline 
 \hline

\end{tabular}}
\label{table_full_no}
\end{table*}

\paragraph{\textbf{Implementation Details}}
Our framework is implemented in PyTorch and trained on Quadro RTX 8000 GPUs. We trained the restoration module for 150k iterations with a batch size of 16. For the generative prior, we employ Stable Diffusion 2.1-base and fine-tune the controllable diffusion module for 205k iterations. The AdamW optimizer is utilized~\cite{adamW} with a learning rate of 0.0001. For inference, we utilize spaced DDPM sampling~\cite{nichol2021improved} with 50 timesteps. Only the low-resolution images (128 × 128 pixels) are input into the trained generative models to assess their performance. The inference process is illustrated in Fig~\ref{infer}.

Additionally, we test the capability of our Histo-Diffusion model to generate images at multiple resolutions. Due to the flexibility of Stable Diffusion's UNet, it can handle input images of any size and perform upscaling to any desired size. This capability allows for multi-scale generation for input images of varying sizes. Patches of low-resolution (LR) images at $5\times$, $10\times$, and $20\times$ magnifications with different image sizes are fed into the trained diffusion model, which then generates corresponding super-resolution (SR) images. This aspect of our research evaluates the model’s effectiveness in multi-resolution image generation, demonstrating its potential utility in diverse clinical scenarios.

\paragraph{\textbf{Evaluation Metrics}} 
We employ commonly used metrics such as PSNR (Peak Signal-to-Noise Ratio), SSIM (Structural Similarity Index Measure), LPIPS (Learned Perceptual Image Patch Similarity)~\cite{lpips}, and ShiftTolerant-LPIPS~\cite{shift} to assess the fidelity of the generated super-resolution images $I_{SR}$ in comparison to the ground truth high-resolution images $I_{HR}$. We also include non-reference metrics such as CLIP-IQA~\cite{clipiqa}, MUSIQ~\cite{musiq}, NIQE~\cite{niqe}, Brisque~\cite{brisque}, and NRQM~\cite{nrqm} to evaluate the realism of all produced images.
Additionally, beyond the commonly used metrics in natural image analysis, we apply specialized full-reference and no-reference metrics specifically designed for the histopathology image super-resolution task. Traditional full-reference metrics such as PSNR and SSIM may not adequately evaluate the complex micro-environment and structural information in histopathology images, potentially leading to misleading results, as demonstrated in Fig~\ref{psnr}. No-reference metrics, typically trained on natural images, lack sufficient exposure to the texture information found in histopathology images, where such details are critical for diagnosis and clinical needs. To address these limitations, we evaluate the super-resolution images using our proposed metrics. These include calculating $L1_{intensity}$ and $L1_{texture}$, as well as assessing blur levels using $CLIP-IQA (boxblur)$ and $CLIP-IQA (Gaussian)$, providing a targeted evaluation of the super-resolution images in the context of histopathology image super-resolution task.

\begin{table*}[t]\small
\centering
\caption{ Comparisons of with and without stage 1 restoration module. The best and worst results are highlighted in \textbf{bold} and \textit{textit}. }
\label{table1}
\resizebox{0.98\textwidth}{!}{%
\begin{tabular}{c|c| c|c c c c c c c c c c |cc c c c}
\hline
\multirow{2}{*}{Datasets} & \multirow{2}{*}{Iterations} & \multirow{2}{*}{Degradation}  & \multicolumn{9}{c}{Metrics} & \multicolumn{4}{c}{Our Metrics}\\\cline{4-17} 
  & &  & & PSNR$\uparrow$ & SSIM$\uparrow$ &LPIPS$\downarrow$& ST-LPIPS$\downarrow$ & CLIP-IQA$\uparrow$ & MUSIQ$\uparrow$ &NIQE$\downarrow$ & BRISQUE$\downarrow$ & NRQM$\uparrow$& $L1_{texture}\downarrow$ & $L1_{intensity}\downarrow$ & CLIP-IQA (boxblur)$\uparrow$ & CLIP-IQA (Gaussian) $\uparrow$\\\hline
\hline
\multirow{2}{*}{PRAD-100} 
% & \multirow{4}{*}{60k} &Real-ESRGAN without stage 1&  &24.70&0.5798&0.2417&0.2194& 0.4808&47.41&4.79&24.81&6.55 &255.63 &13.58 & & \\ 
% &&Real-ESRGAN with stage 1&  &24.67&0.5791&0.2415&0.2160&0.4769&47.51&4.92&25.39&6.66&245.92&13.15 \\
% &&Codeformer without stage1&&24.75&0.5904&0.2369&0.2092&0.4712&46.44&4.69&26.93&6.57&235.52&12.30& &  \\ 
% &&Codeformer with stage 1&&24.72&0.5880&0.2362&0.2097&0.4454&44.57&4.88&29.65&6.46&\textbf{228.80}&\textbf{12.03}& &  \\ 
%  \cline{2-17}
%  & \multirow{4}{*}{100k} &Real-ESRGAN without stage1&  &24.71&0.5802&0.2343&0.2079&0.4771&47.63&4.72&27.26&6.85&261.22 &14.04&\\ 
%  &&Real-ESRGAN with stage1&  &\textit{24.44}& \textit{0.5629} &\textit{0.2453} & 0.2138 &\textbf{0.4828} &\textbf{48.61}& \textbf{4.69}&\textbf{25.90}&\textbf{6.95}&270.34&14.80&\\ 
% &&Codeformer without stage1&  &25.19&0.6083&0.2384&0.2243&0.4361&42.65&5.03&30.47&6.07& 233.99&12.25&\\ 
%  &&Codeformer with stage1&  &24.89&0.6014&0.2280&\textbf{0.2022}&\textit{0.4365}&43.82&4.99&29.70&6.25&\textbf{209.50}&\textbf{11.46}&\\ 
%  \cline{2-17}
 & \multirow{2}{*}{160k} 
 % &Real-ESRGAN without stage1&  &24.70&0.5798&0.2299&0.2026&0.4813&47.88&4.69&26.96&6.79&236.14&12.86&\\ 
 % &&Real-ESRGAN with stage1&  &24.84& 0.5797 &0.2358 & 0.2156 &0.4731&47.31& 4.71&26.60&6.87&266.46&14.20\\ 
 &without stage1&  &\textbf{25.15}&0.6043&0.2368& 0.2216& 0.4378&43.39&\textbf{4.77}&\textbf{29.12}&\textbf{6.26}&241.85&12.47&0.9532&\textbf{0.9540}\\ 
&&with stage1&  &25.03&\textbf{0.6081}&\textbf{0.2287}&\textbf{0.2089}&\textbf{0.4439}&\textbf{43.49}&5.08&30.93&6.16&\textbf{213.36}&\textbf{11.34}&\textbf{0.9541}&0.9529\\ 
\cline{2-13}
% & \multirow{2}{*}{205k} &Real-ESRGAN&  &24.68&0.5741&0.2381&0.2126&0.4732&48.48&4.70&26.86&6.90\\ 
%  &&Codeformer&  &24.86&0.5947&\textbf{0.2279}&0.2079&0.4559&45.00&4.95&\textit{31.39}&6.34\\ 
 \hline \hline

\multirow{2}{*}{LUAD-100}
% & \multirow{4}{*}{60k}  &Real-ESRGAN without stage1& &25.42&0.6539&0.2443&0.2279&0.4178&44.33&5.19&30.17&5.30&215.91&14.83& & \\ 
% &&Real-ESRGAN with stage1&&24.41&0.6085&\textit{0.2972}&\textit{0.2497}&\textit{0.4042}&43.01&5.07&27.77&5.45&251.73&17.05& & \\
% & &Codeformer without stage1& &26.04&0.6878&0.2186&0.2073&0.4425&43.85&5.45&32.98&4.95&\textbf{187.61}&\textbf{12.46}& & \\ 
% & &Codeformer with stage1& &25.79&0.6629&0.2674&0.2313&0.4134&\textit{41.50}&5.24&\textit{34.57}&4.96&188.37&13.17& & \\ 
% \cline{2-17}
% & \multirow{4}{*}{100k}  &Real-ESRGAN without stage1& &25.24&0.6465&0.2559&0.2130&0.4302&45.47&4.84&28.48&5.88& 228.60&15.33&\\ 
% &&Real-ESRGAN with stage1& &\textit{23.90}&\textit{0.5866}&0.2968&0.2472&0.4291&44.52&4.93&\textbf{26.06}&5.78&281.73&18.57&\\ 
% & &Codeformer without stage1&  &27.01&0.7181&0.2307&0.2429&0.3893&38.92&5.69&36.16&4.32& 193.67 &\textbf{12.61}&\\
% & &Codeformer with stage1&  &26.03&0.6785&0.2606&\textbf{0.2249}&0.4181&42.41&5.26&32.41&4.93&\textbf{182.38}&12.86&\\ 
%  \cline{2-17}
&\multirow{2}{*}{160k} 
% &Real-ESRGAN without stage1& &25.31&0.6469&0.2568&0.2121&0.4468&46.16 &4.80&28.43
% &5.81&211.31&14.14\\ 
% &&Real-ESRGAN with stage1& &24.39&0.6013&0.2922&0.2368&0.4186&44.20&\textbf{4.78}&27.62&6.04&267.25&17.17\\
 &without stage1&  &\textbf{26.65}&\textbf{0.7041}&\textbf{0.2240}&\textbf{0.2213}& 0.4124&41.48 &\textbf{5.26}&34.00
&\textbf{4.85}&191.67&\textbf{12.46}&0.9370&0.9457\\ 
& &with stage1&  &26.48&0.6935&0.2576&0.2290&\textbf{0.4133}&\textbf{41.61}&5.33&\textbf{33.08}&4.81&\textbf{176.38}&12.64&\textbf{0.9478}&\textbf{0.9519}\\ 
 \cline{2-13} 
% &\multirow{2}{*}{205k} &Real-ESRGAN& &24.08&0.5902&0.2901&0.2307&0.4202&\textbf{45.77}&4.95&28.08&\textbf{6.09}\\ 
% & &Codeformer&  &25.92&0.6666&\textbf{0.2526}&0.2250&\textbf{0.4309}&43.22&5.12&33.14&5.15\\ 
 \hline \hline

\multirow{2}{*}{GBM-100} 
% &\multirow{4}{*}{60k}  &Real-ESRGAN without stage1&  &22.79&0.4835&0.3031&0.3977&0.4863&49.62&5.34&25.13&6.07& 449.15& 24.12 & &\\ 
% &&Real-ESRGAN with stage1&  &22.68&0.4768&0.3035&\textit{0.3826}&0.5027&51.58&5.26&25.48&6.17&388.25 & 22.00& &\\
% & &Codeformer without stage1 &&23.25&0.5210&0.2837&0.3644&0.5018&50.50&5.16&27.29&5.94&\textbf{287.96}&16.30& &  \\
% & &Codeformer with stage1 &&23.04&0.5087&0.2888&0.3675&0.4819&47.96&5.19&28.78&5.78&293.50&\textbf{15.01}& &  \\ \cline{2-17} 
% & \multirow{4}{*}{100k}  &Real-ESRGAN without stage1&  &22.79&0.4855&0.2975&0.3752&0.5025&50.99&5.09&25.74&6.45& 412.50&22.14&\\ 
% &&Real-ESRGAN with stage1&  &\textit{22.10}&\textit{0.4425}&\textit{0.3059}&0.3784&0.5115&52.53&4.92&21.70&6.70&492.34&26.22& \\ 
% & &Codeformer without stage1&  &23.70&0.5471&0.3016&0.4039& 0.4639&45.42&5.74&32.82&5.01&277.43 &15.26& \\
% & &Codeformer with stage1&  &23.34&0.5357&0.2881&0.3628&0.4650&46.09&5.53&31.70&5.42&\textbf{231.49}&\textbf{12.51}&\\ \cline{2-17}
&\multirow{2}{*}{160k} 
% &Real-ESRGAN without stage1&  &22.55&0.4680&0.2816&0.3573&0.5153& 52.65&4.69&20.83&6.83&395.48&22.52 \\ 
%  & &Real-ESRGAN with stage1&  &22.55&0.4617&0.2921&0.3784&0.5095&51.47&\textbf{4.53}&\textbf{20.84}&\textbf{6.93}&448.60&24.94\\ 
 &without stage1&  &\textbf{23.49}&\textbf{0.5292}&0.2927&0.3790&0.4653&46.49&5.22&29.30&5.44&319.73&17.06&\textbf{0.9566}&\textbf{0.9423} \\
& &with stage1&  &23.25&0.5217&\textbf{0.2822}&\textbf{0.3642}&\textbf{0.4767}&\textbf{47.39}&\textbf{5.05}&\textbf{28.36}&\textbf{5.85}&\textbf{272.71}&\textbf{14.06}&0.9510&0.9361\\ \cline{2-13}
% &\multirow{2}{*}{205k} &Real-ESRGAN&  &22.41&0.4576&0.2988&0.3762&\textbf{0.5126}&\textbf{53.38}&4.77&21.75&6.74 \\ 
% & &Codeformer&  &22.88&0.4950&0.2894&0.3663&0.4862&49.13&4.94&27.52&5.96\\
\hline\hline
\end{tabular}}
\label{table_ablation_wostage1}
\end{table*}

\begin{figure*}[!t]
\centering
\includegraphics[scale=.58]{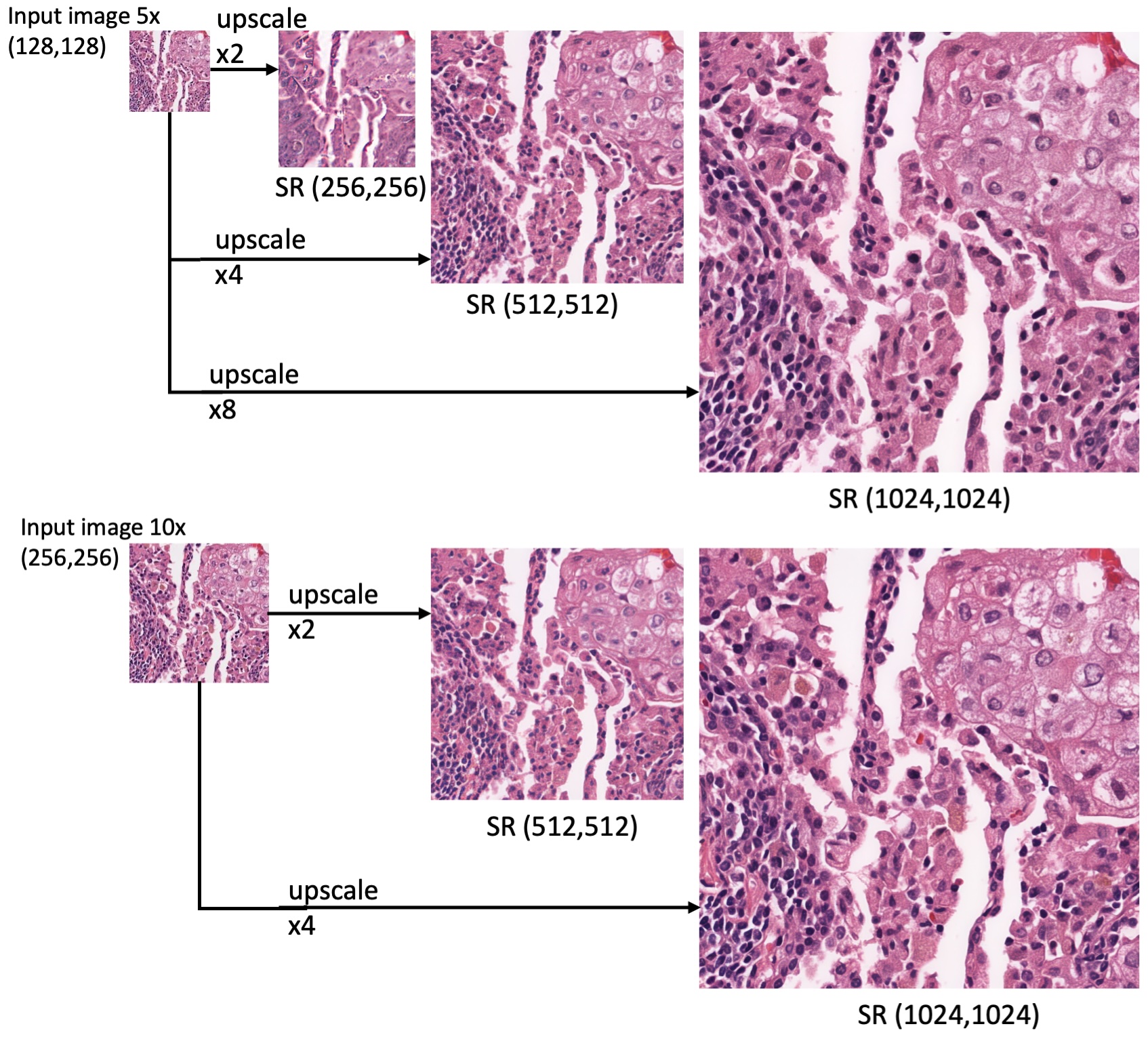}
\caption{Histo-Diffusion supports multi-scale super-resolution image generation, handling various input image sizes and different upscaling factors. For example, when an input image at 5x magnification with dimensions (128, 128) is provided, Histo-Diffusion can upscale it by factors of 2x, 4x, and 8x. It can also accommodate different image sizes, as shown in the second row, where an input image at 10x magnification with dimensions (256, 256) undergoes similar upscaling. This flexibility makes Histo-Diffusion highly adaptable for diverse super-resolution tasks in digital pathology.}
\label{mutli_resolution}
\end{figure*}

\begin{figure*}[!t]
\centering
\includegraphics[scale=.55]{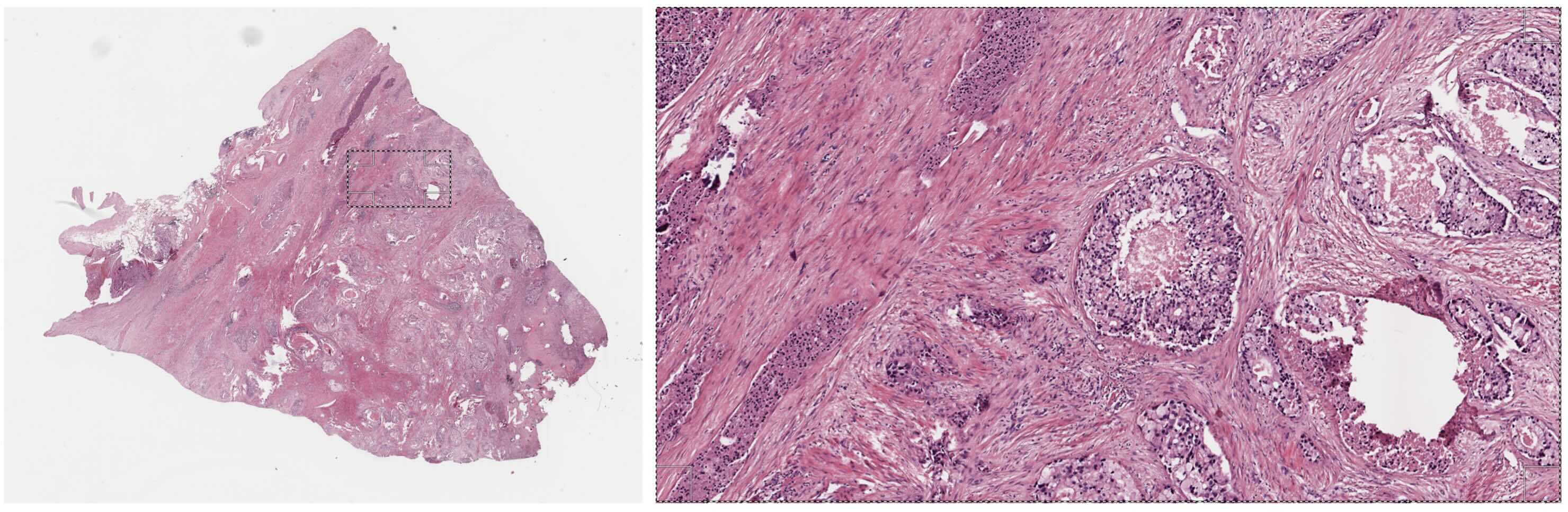}
\caption{A super-resolution image generated at the whole slide level. We input low-resolution images at 5x magnification into our model, which then upscales them by $\times8$ to generate super-resolution patches. The image on the left shows the super-resolution WSI. We've zoomed into a small area within the rectangular box to highlight finer details, as shown in the image on the right. For more examples and WSI-level super resolution images, please visit
\href
  {https://drive.google.com/drive/folders/1keMUeKYULDttIwgdCH68I7ipHCvjQ3HK?usp=sharing}
  {more wsi level examples.}
}
\label{wsi}
\end{figure*}

% \begin{figure*}[!t]
% \centering
% \includegraphics[scale=.58]{figs_lowres/tsne2.jpg}
% \caption{UNI Embedding similarity comparision. We plot convex hull for each pair (SR embedding, HR embedding). Our method's embeddings show the closest overlap to the HR embeddings.}
% \label{tsne}
% \end{figure*}

\subsection{Comparisons with State-of-the-Art Methods}
We evaluate the performance of our proposed Histo-Diffusion method against various state-of-the-art methods including BSRGAN~\cite{bsrgan}, Real-ESRGAN~\cite{realesrgan}, and SwinIR-GAN~\cite{swinir}, utilizing datasets including PRAD-100, PRAD-2000, LUAD-100, and GBM-100. We conduct a detailed examination of both quantitative and qualitative outcomes for these methods across the datasets. Our analysis includes comparisons based on image quality metrics, visual clarity, and their relevance to real-world diagnostic scenarios, providing a comprehensive evaluation of each model's capability to enhance image resolution effectively.

\paragraph{\textbf{Quantitative results}} Low-resolution images with dimensions $128\times128$ are fed into various models to produce super-resolution (SR) images. As detailed in Table~\ref{table_6metric}, our Histo-Diffusion model surpasses state-of-the-art models in perceptual quality, as gauged by ShiftTolerant-LPIPS~\cite{shift}—a recent and robust full-reference image quality assessment (IQA) metric. This metric offers a more precise evaluation than conventional metrics such as PSNR, SSIM, and LPIPS. It also yields comparable results in LPIPS, which assesses the perceptual similarity between images, indicating that our generated SR images bear a closer resemblance to high-resolution ground truth images than those produced by other methods. In terms of no-reference metrics, our model excels in the evaluations of MUSIQ, NIQE, Brisque, and NRQM, suggesting that our generated images not only appear higher in quality but also more realistic compared to those generated by GAN-based methods. These outcomes affirm that the Histo-Diffusion model is adept at generating histopathology images with enhanced realism, outperforming existing SOTA models, particularly those utilizing GANs.

Furthermore, we assess these generated super-resolution images using our proposed full-reference and no-reference metrics, with results shown in Table~\ref{table_full_no}. These results show that our method achieves the closest similarity to high-resolution ground truth images in terms of intensity and texture, highlighting that our approach can closely mimic the texture and intensity levels of the ground truth images. These characteristics, particularly the nuclei position, texture, and intensity properties, are vital for clinical and diagnostic purposes. It is crucial that these generated super-resolution images exhibit properties similar to high-resolution ground truth images in actual nuclear positions. 

Additionally, our generated super-resolution images exhibit sharper details and less blurriness as depicted in Fig~\ref{teaser_0611}. Additional examples of generated super-resolution images can be found in the supplementary materials. Our images also display more accurate color textures compared to GAN-based methods, which sometimes produce inconsistent stain normalization colors in the generated images because they tend to overfit to the specific distribution of training data, whereas diffusion models inherently incorporate noise and iterative refinement, making them more robust to variations in stain and other inconsistencies. Moreover, our images show finer details that align closely with those in the ground truth images, demonstrating that our methods can preserve more details, learn structural information, and produce clearer details, making these images appear more realistic compared to other SOTA methods.

% \paragraph{\textbf{Nuclear segmentation evaluation}} We also explores the downstream applications of super-resolution (SR) images through nuclear segmentation tasks. HoverNet, a mainstream model in nuclear segmentation pre-trained using CoNsep~\cite{hover} (a colorectal nuclear segmentation and phenotypes dataset), is employed to extract nuclear segmentation from both the generated SR images and the ground truth high-resolution (HR) images. We compare all segmentation results from the SR images against those from the HR images, focusing on evaluating the texture and intensity properties at the nuclear level across all images in the test set. Our Uni-Diffusion model excels in achieving the lowest L1 distance for texture and intensity evaluation. This superior performance indicates that our method can generate SR images that closely resemble HR images in terms of texture and intensity, a crucial aspect in the diagnosis using whole slide images.

\paragraph{\textbf{Multiple Upscale Super-Resolution Image Generation with Histo-Diffusion}}

% Unlike previous GAN-based methods that are limited to a single upscaling model, Histo-Diffusion introduces a versatile framework capable of generating super-resolution images at various upscales. This is made possible by Stable Diffusion's UNet, which can handle any size of latent. Our method's adaptability allows it to process various input image sizes and different upscales. 
We present multiple upscaled super-resolution images in Figure~\ref{mutli_resolution}. Our model is capable of handling various input image sizes and different upscaling factors.. The generated super-resolution images retain texture and intensity fidelity across various input image sizes and scaling factors. This flexibility enables the use of a single trained model for multi-resolution SR image generation tasks in histopathology, significantly reducing training time and enhancing efficiency. This capability paves the way for clinical utilization, especially in scenarios where high-resolution images at 40x are scarce.

% This model's flexibility presents a considerable advantage over current state-of-the-art methods, which often require fixed-scale enhancements. GAN architectures, typically confined to a specific upscale factor, usually need modifications for multi-scale super-resolution image generation. In contrast, Histo-Diffusion enables flexible, multi-scale super-resolution image generation without additional model training. Results depicted in Figure \ref{mutli_resolution} confirm that the super-resolution images retain texture and intensity fidelity across various magnifications from 5x to 40x. This demonstrates the generated images' high realism and structural integrity by Histo-Diffusion, providing a viable solution when only low-resolution images are available. These super-resolution images can subsequently serve as high-resolution equivalents for clinical diagnosis.

\paragraph{\textbf{Whole Slide Image Level Super-Resolution}} The Histo-Diffusion model's adaptability allows for the generation of super-resolution WSIs from lower magnification WSIs. As depicted in Figure \ref{wsi}, the model has effectively produced a super-resolution WSI-level image from low-resolution images at 5x. This feature is particularly advantageous for researchers who do not have access to high-resolution scanning equipment. By upgrading low-magnification images to higher resolutions, Histo-Diffusion can not only support the diagnostic process but also enhances the performance of various downstream tasks that require higher image quality. This capability ensures that detailed cellular and tissue structures are preserved and enhanced, facilitating more accurate analyses and interpretations in medical research and clinical settings.

\subsection{Ablation study}

We also evaluate the effectiveness of the restoration module, with results shown in Table~\ref{table_ablation_wostage1}. Without the stage 1 restoration module, the resized control image is directly fed into the controllable diffusion module. The results indicate that including stage 1 yields better performance in PRAD-100 and GBM-100, specifically in texture and intensity similarity compared to high-resolution ground truth images. Additionally, the CLIP-IQA scores for these two datasets are comparable. For LUAD-100, our method demonstrates improved texture similarity, CLIP-IQA scores, and comparable intensity similarity results. This suggests that the restoration module enhances image quality and generates more realistic super-resolution images that closely match high-resolution ground truth images. Further results from the ablation study are provided in the supplementary materials.

\subsection{Downstream tasks}
To further assess the differences between the generated super-resolution images and ground truth high-resolution images to see whether these SR images can be leveraged for downstream tasks, we evaluate the similarity between SR embeddings and HR embeddings. For PRAD-100, we extract the UNI~\cite{uni} and PLIP~\cite{plip} embeddings for the SR images and the corresponding HR images. Then we calculate the cosine similarity between the SR embeddings and HR embeddings to measure the \textbf{Embedding Similarity}. The results have been shown in Table~\ref{embedding_similarity}. We can see that our generated SR images' embeddings show highest similarity scores across different foundation models, which suggests that using our generated SR images can yeild similar embeddings to those of HR images. This similarity demonstrates the potential of using our generated images for further downstream tasks such as classification and segmentation.

\begin{table}[t]
\small
\centering
\caption{Embedding Similarities for PRAD-100 dataset}
\resizebox{0.45\textwidth}{!}{%
\begin{tabular}{c| c c c c c}
\hline
 & Bicubic & BSRGAN & Real-ESRGAN & SwinIR-GAN & Ours\\\hline
UNI Embedding Similarity & 0.7139 & 0.6885 & 0.7317 & 0.7818& \textbf{0.8348} \\\hline
PLIP Embedding Similarity & 0.9058 & 0.9394 & 0.9383 & 0.9457& \textbf{0.9662} \\\hline\hline

\end{tabular}}
\label{embedding_similarity}
\end{table}

\section{Conclusions}

Histo-Diffusion effectively addresses the limitations of traditional super-resolution techniques in computational pathology. 
% Designed to generate super-resolution images with high fidelity and realism, Histo-Diffusion is particularly suited for histopathology image field. 
Our comprehensive evaluation methodology, supported by two specially curated histopathology image quality assessment datasets, ensures a thorough quality assessment using both full-reference and no-reference metrics.

One of the key strengths of Histo-Diffusion is its ability to outperform GAN-based models, offering a versatile and adaptable solution capable of handling multi-resolution generation across varied input sizes. This approach overcomes the constraints of conventional state-of-the-art methods, which are typically limited to fixed upscale factors. Consequently, Histo-Diffusion provides a versatile solution for histopathology image super-resolution task.

The proposed evaluation metrics demonstrate that the generated super-resolution images closely align with high-resolution ground truths. This alignment makes the images well-suited for critical downstream tasks, such as nuclear segmentation and diagnostic support. This capability makes it a valuable tool when the high resolution images are not available.

Despite its advantages, Histo-Diffusion has certain limitations. Currently, our model is trained on a single cancer type (TCGA-PRAD). Despite our single cancer-type trained model showing promising performance across multiple cancer types, we believe that training on more cancer types in the future and developing a foundational super-resolution model could better capture the diversity of pan-cancer. Additionally, while our evaluation metrics are comprehensive, further validation on larger and more diverse datasets would be beneficial to fully establish the model's robustness and reliability.

To address the current limitations, future work will focus on expanding the training dataset to include patches from all TCGA cancer types. This expansion aims to develop a foundation model for histopathology images, enhancing the diversity and applicability of Histo-Diffusion across a broader range of clinical scenarios. Additionally, we plan to further refine our evaluation metrics and explore more advanced no-reference quality assessment techniques~\cite{maniqa,reiqa} to better capture the nuances of histopathology images.

In summary, Histo-Diffusion represents a major breakthrough in digital pathology, offering a robust, efficient, and adaptable solution for super-resolution image generation and evaluation. Its capability to produce high-quality images suitable for clinical use, combined with its comprehensive evaluation methodology, showcases its potential to become an essential tool in the computational pthology workflow.

\section{Acknowledgements}
Reported research was supported by the OVPR seed grant and ProFund grant at Stony Brook University, NIH 1R01CA297843-01, and  NIH 1R21CA258493-01A1. The content is solely the responsibility of the authors and does not necessarily represent the official views of the National Institutes of Health. 

\section{Declaration of generative AI and AI-assisted technologies in the writing process}
During the preparation of this work the authors used ChatGPT to correct grammatical errors. The authors reviewed and edited the content as needed and take full responsibility for the content of the publication.

\bibliographystyle{model2-names.bst}\biboptions{authoryear}
\clearpage
\bibliography{refs}
\clearpage
\section*{Supplementary Material}

\paragraph{\textbf{Visual comparisons}} 
\begin{figure*}[!t]
\centering
\includegraphics[scale=.06]{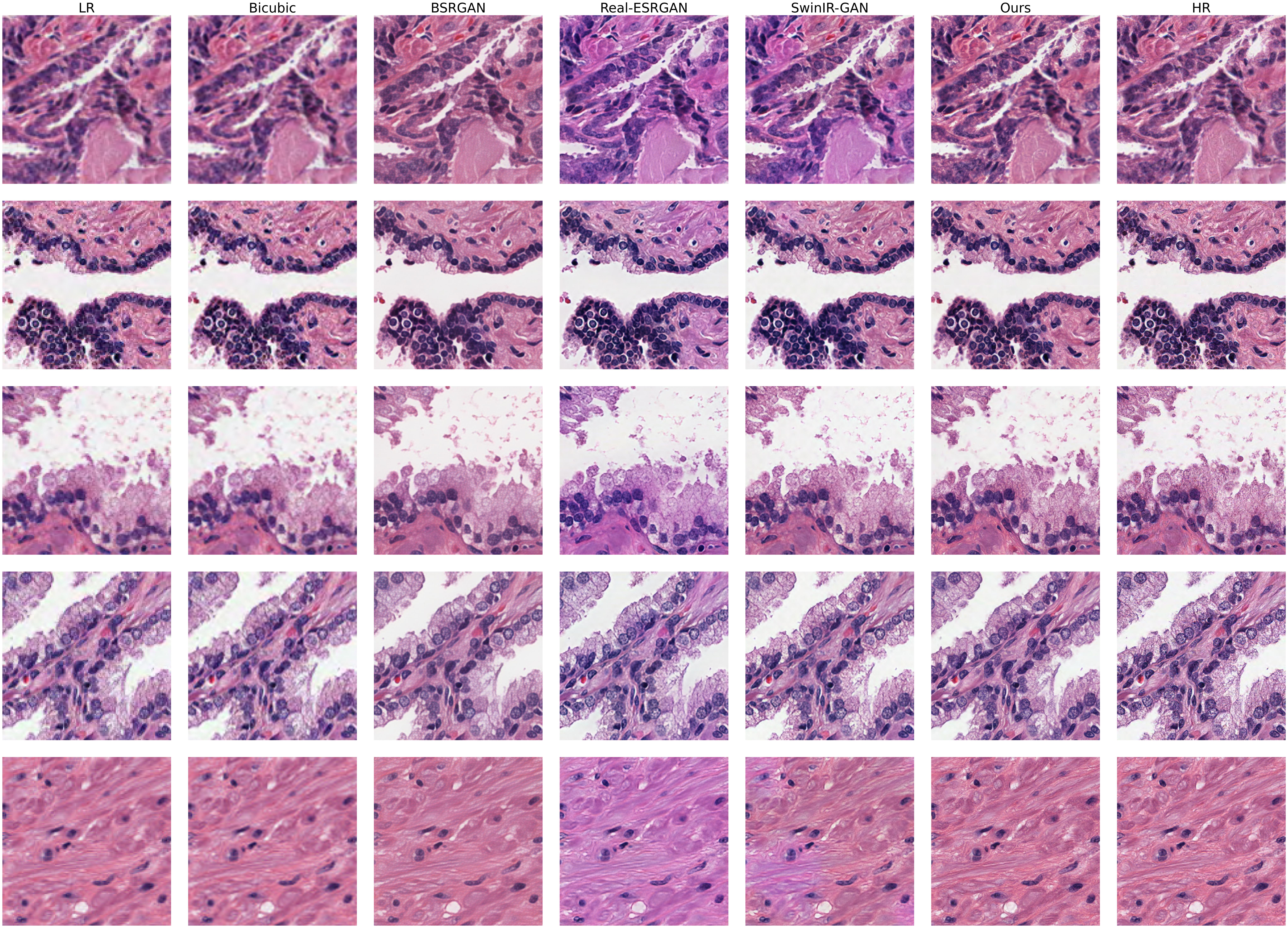}
\caption{ Visual comparisons on PRAD-100 samples. Please zoom in for more details.}
\label{teaser_prad100}
\end{figure*}

\begin{figure*}[!t]
\centering
\includegraphics[scale=.05]{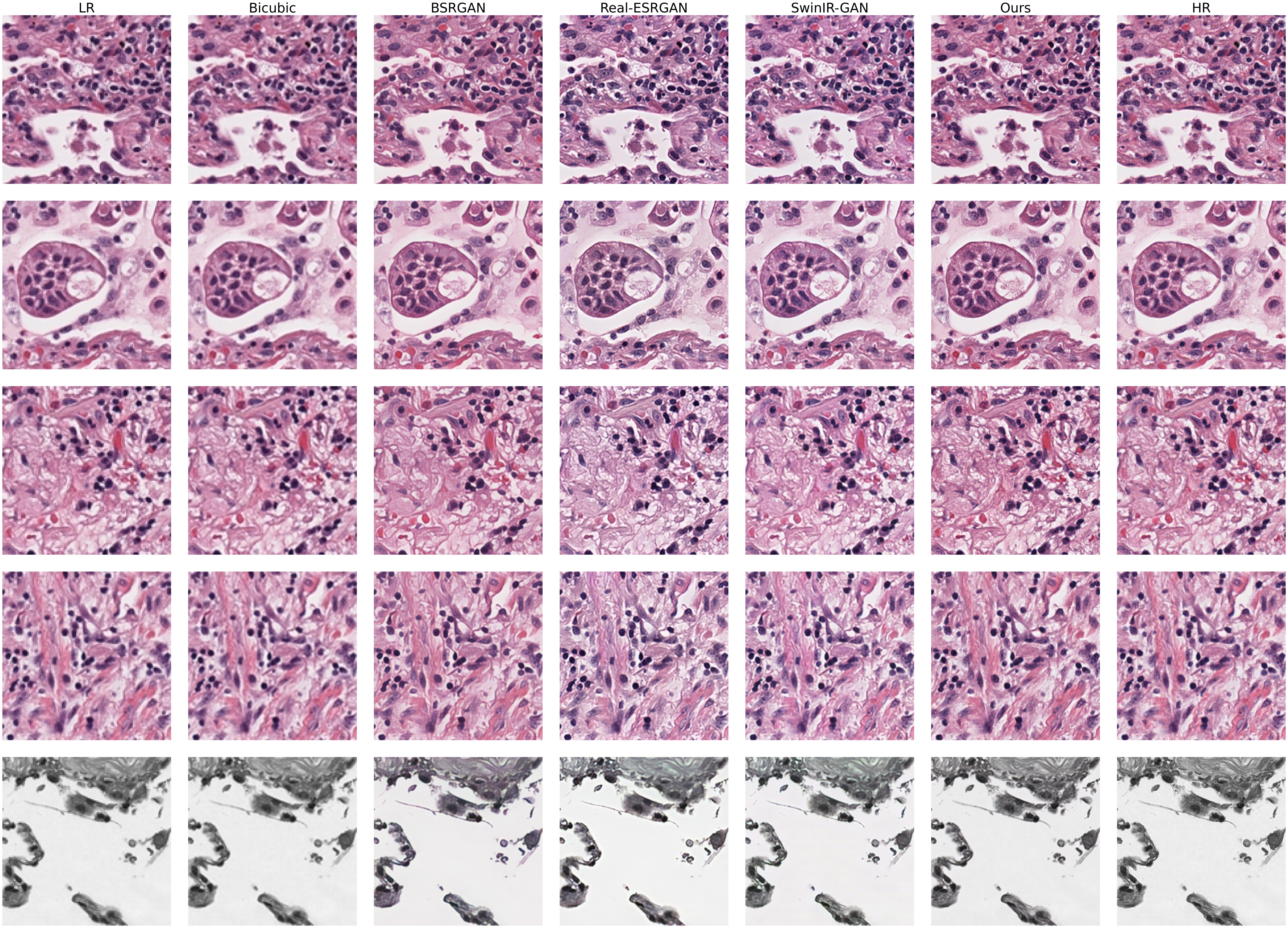}
\caption{ Visual comparisons on LUAD-100 samples. Please zoom in for more details.}
\label{teaser_luad100}
\end{figure*}

\begin{figure*}[!t]
\centering
\includegraphics[scale=.05]{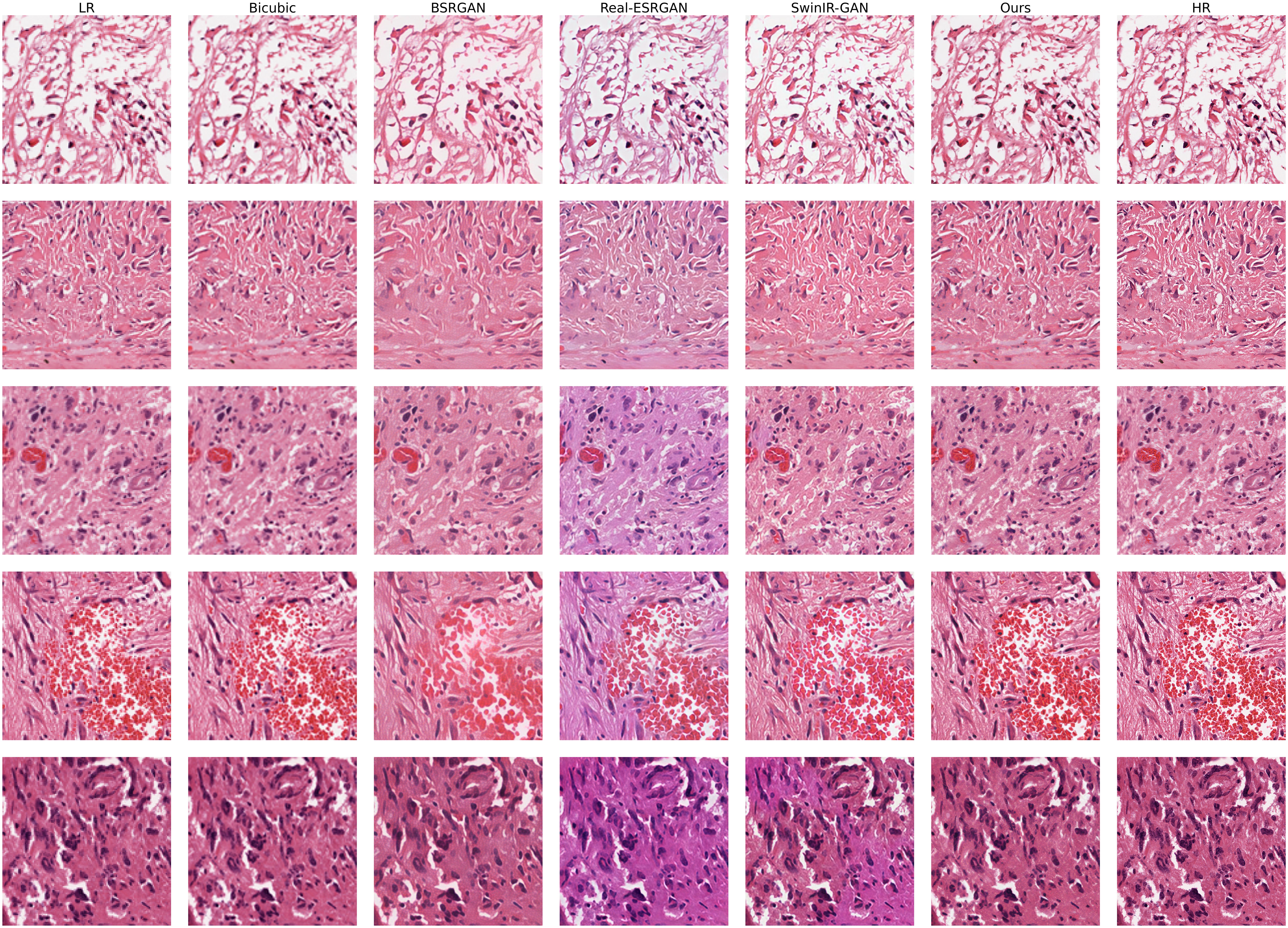}
\caption{ Visual comparisons on GBM-100 samples. Please zoom in for more details.}
\label{teaser_gbm100}
\end{figure*}
We have included additional visual comparison examples of generated super-resolution images using both GAN-based methods and our diffusion-based method, as shown in Figures \ref{teaser_prad100}, \ref{teaser_luad100}, and \ref{teaser_gbm100}. Our diffusion-based approach generates super-resolution images with sharper details and more stable color staining that closely resemble high-resolution images. Furthermore, our method maintains texture information similar to the high-resolution ground truth images. Please zoom in to see the details more clearly.

\paragraph{\textbf{Dataset distribution.}} We have provide the image names and their corresponding WSI ids in google drive. Please visit
\href
  {https://drive.google.com/drive/folders/1keMUeKYULDttIwgdCH68I7ipHCvjQ3HK?usp=sharing}
  {WSI IDs.} for more information. There is no overlapping between training set and test set.

\paragraph{\textbf{Analysis of Histo-Diffusion}}
We also explore how different types of degradation impact the Histo-Diffusion model, which consists of a stage 1 SwinIR restoration module and a stage 2 controllable diffusion module. Both stages maintain the same settings, with the primary difference being the type of degradation applied in stage 1. Here, HR images are intentionally degraded to simulate real-world histopathological conditions, resulting in degraded LR images. These LR images are then restored using the SwinIR module to produce $I_{RM}$.

For the degradation process, we utilize the degradation type from CodeFormer~\cite{codeformer}, known for its ability to enhance detail in specific face restoration tasks where detail retention is critical. Additionally, we compare with degradation type from Real-ESRGAN~\cite{realesrgan} here, which involves a second-order degradation and offers broader applicability across various image super-resolution tasks. Both degradation types effectively handle real-world noise, blur, and compression artifacts in natural images, and we analyze their effectiveness on digital pathology images.

Moreover, we assess how training iterations influence the performance of the generated super-resolution images. Considering that prolonged training durations and high computational demands are significant challenges of diffusion models, we monitor performance across various training iterations. Additionally, to determine if extended training could lead to overfitting—a frequent issue with GANs—we evaluate the model's performance at various training milestones (60k, 100k, 160k, and 205k iterations) across different datasets: PRAD-100, LUAD-100, and GBM-100. This evaluation allows us to observe changes in performance and assess the potential for overfitting as training progresses.

The results presented in Table~\ref{table_ablation_dataset} illustrate our findings. We compared different training iterations and degradation types, revealing that even with the smallest training iteration of 60k, the performance remains competitive. When comparing GAN-based and diffusion-based methods, even the least-performing model in Table~\ref{table_ablation_dataset} and Table~\ref{table_6metric} consistently surpasses GANs in ST-LPIPS, MUSIQ, NIQE, BRISQUE, and NRQM across all three datasets. Additionally, it shows superior LPIPS results for LUAD-100 and GBM-100, indicating that the method generalizes well to different cancer types while maintaining high perceptual similarity. These results demonstrate that our approach can produce high quality super-resolution images that closely resemble high-resolution ground truths. 

In comparing different degradation types, CodeFormer degradation performs better in full-reference IQA (LPIPS and ST-LPIPS), while Real-ESRGAN degradation excels in no-reference IQA (CLIP-IQA, MUSIQ, NIQE, BRISQUE, and NRQM). Both show competitive results compared to GANs, prompting the question of whether degradation simulation, crucial in GAN-based methods for mimicking real-world degradation~\cite{realesrgan}, is as important in diffusion-based super-resolution models.

In the forward phase, a diffusion model progressively adds noise to an image, simulating various real-world degradations. The reverse phase systematically removes the noise, restoring image details and reversing the degradation. This inherent mechanism addresses the degradation mimic problem in histopathology by methodically eliminating various forms of noise and artifacts.

In histopathology, degradation often arises from poor staining, variations in slide preparation, or suboptimal imaging conditions, leading to blurring, noise, and artifacts that obscure critical diagnostic information. Traditional methods struggle to replicate and correct these degradation patterns without matching training data. However, Histo-Diffusion provides a way to inherently address the degradation mimic problem in histopathology by learning to simulate and systematically eliminate noise and artifacts. This makes our diffusion-based super-resolution model highly suitable for enhancing image quality in fields like histopathology, where managing diverse and unpredictable degradations is crucial.
\begin{table*}[t]\small
\centering
\caption{ Comparisons of different image degradations (Real-ESRGAN degradation vs CodeFormer degradation) for different iterations on histopathology image datasets with a 4× upsampling scale using our Histo-Diffusion.  The best and worst results are highlighted in \textbf{bold} and \textit{textit}. }
\label{table1}
\resizebox{0.98\textwidth}{!}{%
\begin{tabular}{c|c| c|c c c c c c c c c c c}
\hline
\multirow{2}{*}{Datasets} & \multirow{2}{*}{Iterations} & \multirow{2}{*}{Degradation}  & \multicolumn{9}{c}{Metrics}\\\cline{4-13} 
  & &  & & PSNR$\uparrow$ & SSIM$\uparrow$ &LPIPS$\downarrow$& ST-LPIPS$\downarrow$ & CLIP-IQA$\uparrow$ & MUSIQ$\uparrow$ &NIQE$\downarrow$ & BRISQUE$\downarrow$ & NRQM$\uparrow$\\\hline
\hline
\multirow{6}{*}{PRAD-100} & \multirow{2}{*}{60k} &Real-ESRGAN&  &24.67&0.5791&0.2415&\textit{0.2160}&0.4769&47.51&4.92&25.39&6.66 \\ 
&&Codeformer&&24.72&0.5880&0.2362&0.2097&0.4454&44.57&4.88&29.65&6.46  \\ 
 \cline{2-13}
 & \multirow{2}{*}{100k} &Real-ESRGAN&  &\textit{24.44}& \textit{0.5629} &\textit{0.2453} & 0.2138 &\textbf{0.4828} &\textbf{48.61}& \textbf{4.69}&\textbf{25.90}&\textbf{6.95}\\ 
 &&Codeformer&  &24.89&0.6014&0.2280&\textbf{0.2022}&\textit{0.4365}&43.82&4.99&29.70&6.25\\ 
 \cline{2-13}
 & \multirow{2}{*}{160k} &Real-ESRGAN&  &24.84& 0.5797 &0.2358 & 0.2156 &0.4731&47.31& 4.71&26.60&6.87\\ 
&&Codeformer&  &\textbf{25.03}&\textbf{0.6081}&0.2287&0.2089&0.4439&\textit{43.49}&\textit{5.08}&30.93&\textit{6.16}\\ 
\cline{2-13}
& \multirow{2}{*}{205k} &Real-ESRGAN&  &24.68&0.5741&0.2381&0.2126&0.4732&48.48&4.70&26.86&6.90\\ 
 &&Codeformer&  &24.86&0.5947&\textbf{0.2279}&0.2079&0.4559&45.00&4.95&\textit{31.39}&6.34\\ 
 \hline \hline

\multirow{6}{*}{LUAD-100}& \multirow{2}{*}{60k}  &Real-ESRGAN& &24.41&0.6085&\textit{0.2972}&\textit{0.2497}&\textit{0.4042}&43.01&5.07&27.77&5.45\\ 
& &Codeformer& &25.79&0.6629&0.2674&0.2313&0.4134&\textit{41.50}&5.24&\textit{34.57}&4.96 \\ 
\cline{2-13}
& \multirow{2}{*}{100k}  &Real-ESRGAN& &\textit{23.90}&\textit{0.5866}&0.2968&0.2472&0.4291&44.52&4.93&\textbf{26.06}&5.78\\ 
& &Codeformer&  &26.03&0.6785&0.2606&\textbf{0.2249}&0.4181&42.41&5.26&32.41&4.93\\ 
 \cline{2-13}
&\multirow{2}{*}{160k} &Real-ESRGAN& &24.39&0.6013&0.2922&0.2368&0.4186&44.20&\textbf{4.78}&27.62&6.04\\ 
& &Codeformer&  &\textbf{26.48}&\textbf{0.6935}&0.2576&0.2290&0.4133&41.61&\textit{5.33}&33.08&\textit{4.81}\\ 
 \cline{2-13} 
&\multirow{2}{*}{205k} &Real-ESRGAN& &24.08&0.5902&0.2901&0.2307&0.4202&\textbf{45.77}&4.95&28.08&\textbf{6.09}\\ 
& &Codeformer&  &25.92&0.6666&\textbf{0.2526}&0.2250&\textbf{0.4309}&43.22&5.12&33.14&5.15\\ 
 \hline \hline

\multirow{6}{*}{GBM-100}& \multirow{2}{*}{60k}  &Real-ESRGAN&  &22.68&0.4768&0.3035&\textit{0.3826}&0.5027&51.58&5.26&25.48&6.17\\ 
& &Codeformer&&23.04&0.5087&0.2888&0.3675&0.4819&47.96&5.19&28.78&5.78  \\ \cline{2-13} 
& \multirow{2}{*}{100k}  &Real-ESRGAN&  &\textit{22.10}&\textit{0.4425}&\textit{0.3059}&0.3784&0.5115&52.53&4.92&21.70&6.70\\ 
& &Codeformer&  &\textbf{23.34}&\textbf{0.5357}&0.2881&\textbf{0.3628}&\textit{0.4650}&\textit{46.09}&\textit{5.53}&\textit{31.70}&\textit{5.42}\\ \cline{2-13}
&\multirow{2}{*}{160k} &Real-ESRGAN&  &22.55&0.4617&0.2921&0.3784&0.5095&51.47&\textbf{4.53}&\textbf{20.84}&\textbf{6.93}\\ 
& &Codeformer&  &23.25&0.5217&\textbf{0.2822}&0.3642&0.4767&47.39&5.05&28.36&5.85\\ \cline{2-13}
&\multirow{2}{*}{205k} &Real-ESRGAN&  &22.41&0.4576&0.2988&0.3762&\textbf{0.5126}&\textbf{53.38}&4.77&21.75&6.74 \\ 
& &Codeformer&  &22.88&0.4950&0.2894&0.3663&0.4862&49.13&4.94&27.52&5.96\\\hline\hline
\end{tabular}}
\label{table_ablation_dataset}
\end{table*}

\end{document}